\documentclass[twocolumn,prx,aps,amsmath,amssymb,longbibliography]{revtex4-1}

\usepackage{graphicx}
\usepackage{dcolumn}
\usepackage{bm}

\begin{document}

\title{Universal Majorana thermoelectric noise}

\author{Sergey Smirnov}
\affiliation{P. N. Lebedev Physical Institute of the Russian Academy of
  Sciences, 119991 Moscow, Russia}
\email{sergej.physik@gmail.com,  ssmirnov@sci.lebedev.ru}

\date{\today}

\begin{abstract}
Thermoelectric phenomena resulting from an interplay between particle flows
induced by electric fields and temperature inhomogeneities are extremely
insightful as a tool providing substantial knowledge about the microscopic
structure of a given system. Tuning, {\it e.g.}, parameters of a nanoscopic
system coupled via tunneling mechanisms to two contacts one may achieve
various situations where the electric current induced by an external bias
voltage competes with the electric current excited by the temperature
difference of the two contacts. Even more exciting physics emerges when the
system's electronic degrees freedom split to form Majorana fermions which make
the thermoelectric dynamics universal. Here we propose revealing this unique
universal signatures of Majorana fermions in strongly nonequilibrium quantum
dots via noise of the thermoelectric transport beyond linear response. It is
demonstrated that whereas mean thermoelectric quantities are only universal at
large bias voltages, the noise of the electric current excited by an external
bias voltage and the temperature difference of the contacts is universal at
any bias voltage. We provide truly universal, {\it i.e.} independent of the
system's parameters, thermoelectric ratios between nonlinear response
coefficients of the noise and mean current at large bias voltages where
experiments may easily be performed to uniquely detect these truly universal
Majorana thermoelectric signatures.
\end{abstract}


\maketitle

\section{Introduction}\label{intro}
Originally proposed \cite{Majorana_1937} in the late 1930s the Majorana
representation \cite{Itzykson_Zuber_1980}, making the Dirac equation real, has
since then always been a challenge for experiments on elementary particles to
reveal, {\it e.g.} via neutrinoless double beta decay, a fundamental particle
representing on equal footing its own antiparticle. Referred to as Majorana
fermions, these are hypothetical spin-1/2 neutral fundamental particles with
neutrinos as possible candidates for massive Majorana particles.

In parallel to Majorana challenges within the particle physics another way to
implement Majorana fermions is to construct condensed matter systems with
quasiparticles being identical to their own antiquasiparticles. This is indeed
possible when a system acquires a finite superconducting order parameter. Here
Dirac fermions may become highly nonlocal in real space, {\it i.e.} they may
split, or fractionalize, into two Majorana fermions localized at the system's
edges. Although these zero energy Majorana bound states are not fundamental
particles as in the case of the particle physics, they are effectively
quasiparticles which are their own antiquasiparticles. Despite their
non-abelian statistics these zero energy Majorana bound states are still
referred to as Majorana fermions. The Kitaev model \cite{Kitaev_2001} is an
example of a one-dimensional condensed matter system hosting two Majorana
fermions at its ends. The topological superconducting phase realized in the
Kitaev model has various \cite{Alicea_2012,Flensberg_2012,Sato_2016} physical
implementations among which systems based on topological insulators
\cite{Fu_2008,Fu_2009} and spin-orbit coupled semiconductors
\cite{Lutchyn_2010,Oreg_2010} are of particular interest.

Once implemented in a condensed matter setup Majorana fermions need an
experimental proof of their existence via unique signatures characteristic of
exclusively these fractionalized quasiparticles. To this end transport
experiments offer a relatively simple way to detect unique Majorana signatures
in a given condensed matter system.

The majority of transport proposals focus on mean electric current often
providing low-energy behavior of the system's electrical conductance. Examples
are given by superconductor-Luttinger liquid junctions \cite{Fidkowski_2012},
Kondo effect in quantum dots side coupled to a topological superconductor
supporting Majorana fermions at its ends \cite{Lee_2013}, driven topological
superconductors \cite{Kundu_2013}, Josephson junctions on surfaces of
three-dimensional topological insulators \cite{Ilan_2014}, Kondo effect in
topological superconductor-quantum dot-normal lead junction \cite{Cheng_2014},
normal metal-superconducting semiconductor-normal metal structures
\cite{Lobos_2015}, spinon-antispinon systems \cite{Wang_2016}, Coulomb
blockaded systems \cite{Lutchyn_2017}, disordered Josephson junctions in
tilted magnetic fields \cite{Huang_2017} and many others.

Transport experiments \cite{Mourik_2012,Albrecht_2016} oriented on
measurements of the mean current, in particular, on its low-energy behavior
are steadily improving their analysis of the zero bias anomaly present in the
electrical conductance from which stronger signatures of Majoranas may be
extracted \cite{Zhang_2017}. Although it is often necessary to perform
additional analysis to disentangle the Majorana physics from possible
interaction induced effects, such as the Kondo effect, or from effects related
to partially-separated Andreev bound states
\cite{Kells_2012,Liu_2017}, recent works \cite{Prada_2017,Clarke_2017} suggest
an alternative proof of the nonlocal nature and topological protection of
Majorana zero modes via local measurements of the zero bias conductance.

To avoid uncertainty and make transport experiments provide more unique
signatures of Majorana fermions one might, in conjunction with measurements of
zero bias conductance, resort to measurements of the electric current
fluctuations. Here there are not as many proposals
\cite{Liu_2015,Liu_2015a,Beenakker_2015,Valentini_2016} for Majorana noise as
for the mean electric current. Again the most attention is paid to the
low-energy behavior of the electric current noise. Because of the
fluctuation-dissipation theorem \cite{Nyquist_1928,Callen_1951,Landau_V} this
equilibrium noise, accessed within the system's linear response, is not
independent of the mean electric current. Beyond linear response the electric
current fluctuates independently of its mean value and thus this
nonequilibrium noise is able to provide alternative Majorana fingerprints
which, what is remarkable, turn out to be of universal nature
\cite{Haim_2015,Smirnov_2017}. In particular, in Ref. \cite{Smirnov_2017} it
has been demonstrated that this universal Majorana noise is characterized by
two effective charges $e^*_l=e/2$ and $e^*_h=3e/2$ at low and high energies,
respectively. It has also been shown that the low-energy Majorana effective
charge $e^*_l=e/2$, whose Majorana nature is additionally confirmed via
Majorana tunneling entropy calculations \cite{Smirnov_2015}, might be
sensitive to thermal fluctuations while the high-energy Majorana effective
charge $e^*_h=3e/2$ is accessed at high bias voltages which protect this
effective charge from thermal noise and, therefore, its measurement may be
performed at relatively high temperatures easily reachable in modern
laboratories. It is important to note that modern experiments have already
reached such a high level at which one can get the effective charge with very
high accuracy as for example in experiments on quantum dots where the
effective charge has been measured in the Kondo regime \cite{Ferrier_2016}.

Another realm of transport experiments is provided by systems hosting Majorana
fermions and possessing temperature inhomogeneities. Here examples include
heat conduction in a Majorana metal, where the heat conduction happens via
Majorana fermions bound to defects \cite{Senthil_2000}, and thermal quantum
Hall effect \cite{Read_2000}, where a net heat current results from a
temperature difference between the edges which support Majorana chiral edge
modes arising for example in a chiral $p$-wave superconductor. The latter case
of dispersive Majorana fermions is particularly attractive for future research
on thermoelectric phenomena in systems where Majorana dispersion relations may
undergo qualitative changes such as, for example, the evolution of a Majorana
conic dispersion relation into a Majorana arc shaped dispersion relation
\cite{Mizushima_2018}. In the context of nanoscopic systems one is usually
interested in setups where a nanoscopic system interacts via tunneling
mechanisms with two normal metals playing the role of the contacts. An
external bias voltage is applied to these contacts which may also have
different temperatures. Here one may easily tune the system's parameters
making the electric current induced by the bias voltage and the electric
current excited by the temperature difference interfere with one another
leading to a highly nontrivial thermoelectric transport
\cite{Dorda_2016,Titvinidze_2017}. When a nanoscopic setup is designed to host
Majorana degrees of freedom, the system's thermoelectric response undergoes
essential changes which might be used to identify Majorana fermions in,
{\it e.g.}, normal metal-quantum dot-Majorana bound states junction
\cite{Leijnse_2014}, Majorana-side-coupled quantum dots
\cite{Lopez_2014,Khim_2015}, Majorana bound states system coupled to two
normal leads \cite{Ramos-Andrade_2016}.

Up to now Majorana thermoelectric transport has been investigated via
analysis of the mean electric current often restricted to its low-energy
behavior to obtain linear conductances. Fluctuations of the electric current
excited by both the bias voltage and temperature difference have not been
addressed. At the same time, as discussed above, fluctuations provide much
more unique fingerprints of a given system. Therefore, thermoelectric noise
is an important and highly conclusive tool to uniquely reveal whether Majorana
fermions are present in a given nanoscopic system.

In this work we explore fluctuations of thermoelectric transport through
quantum dots interacting via tunneling mechanisms with two contacts and one
end of a one-dimensional Kitaev's chain, supporting two Majorana fermions at
its ends. The contacts are normal metals to which an external bias voltage is
applied and in general the temperatures of the two contacts are not
identical. We obtain for this system both the mean electric current and
thermoelectric noise as well as their nonlinear response coefficients in the
transport regime governed by Majorana degrees of freedom. We find 1) that mean
thermoelectric quantities become universal only for large bias voltages while
having non-universal behavior below the energy scale characterizing the
overlap of the Majorana modes; 2) that in contrast to the mean quantities
thermoelectric noise is universal in the whole voltage range when the dynamics
is essentially governed by Majorana bound states; 3) that the differential
thermoelectric noise has a universal two plateau structure with the truly
universal values of the plateaus $(e^3/h)[1+\ln(2)]$ and
$(e^3/h)[1+\ln(2^{1/2})]$ depending on whether the ratio between the bias
voltage and the thermal voltage is less than or greater than one; 4)
analytical universal high-energy asymptotics of the nonlinear response
coefficients of the thermoelectric noise; 5) universal ratios between
nonlinear response coefficients of the thermoelectric noise and mean current;
6) that at large bias voltages these thermoelectric ratios saturate to truly
universal constants independent of the bias voltage; 7) analytical values of
these constants; 8) that these values are protected by high bias voltages,
making them robust against thermal noise and, therefore, these values
represent unique truly universal Majorana thermoelectric signatures which
could be measured in modern experiments even at relatively high temperatures.

The paper is organized as follows. In Section \ref{fld_thr} we describe in
detail an example of a system where thermoelectric transport in presence of
Majorana fermions may be analyzed on the level of both mean quantities and
fluctuations. The description of the system is given in terms of the Keldysh
field integral most convenient to explore different correlation functions in
stationary nonequilibrium which is the case here.  In Section
\ref{un_ma_thrmel_tr} thermoelectric transport is analyzed in terms of
nonlinear response coefficients: first on the level of mean quantities, in
Subsection \ref{res_curr}, and afterwards on the level of fluctuations, in
Subsection \ref{res_noise}. In Section \ref{concl} we summarize the main
results of the paper, draw conclusions and discuss some open issues. Finally,
the appendix provides the main technical steps in the calculation of the
current-current correlator.
\section{Nonequilibrium field-theoretic framework: Keldysh action}\label{fld_thr}
\begin{figure}
\includegraphics[width=8.0 cm]{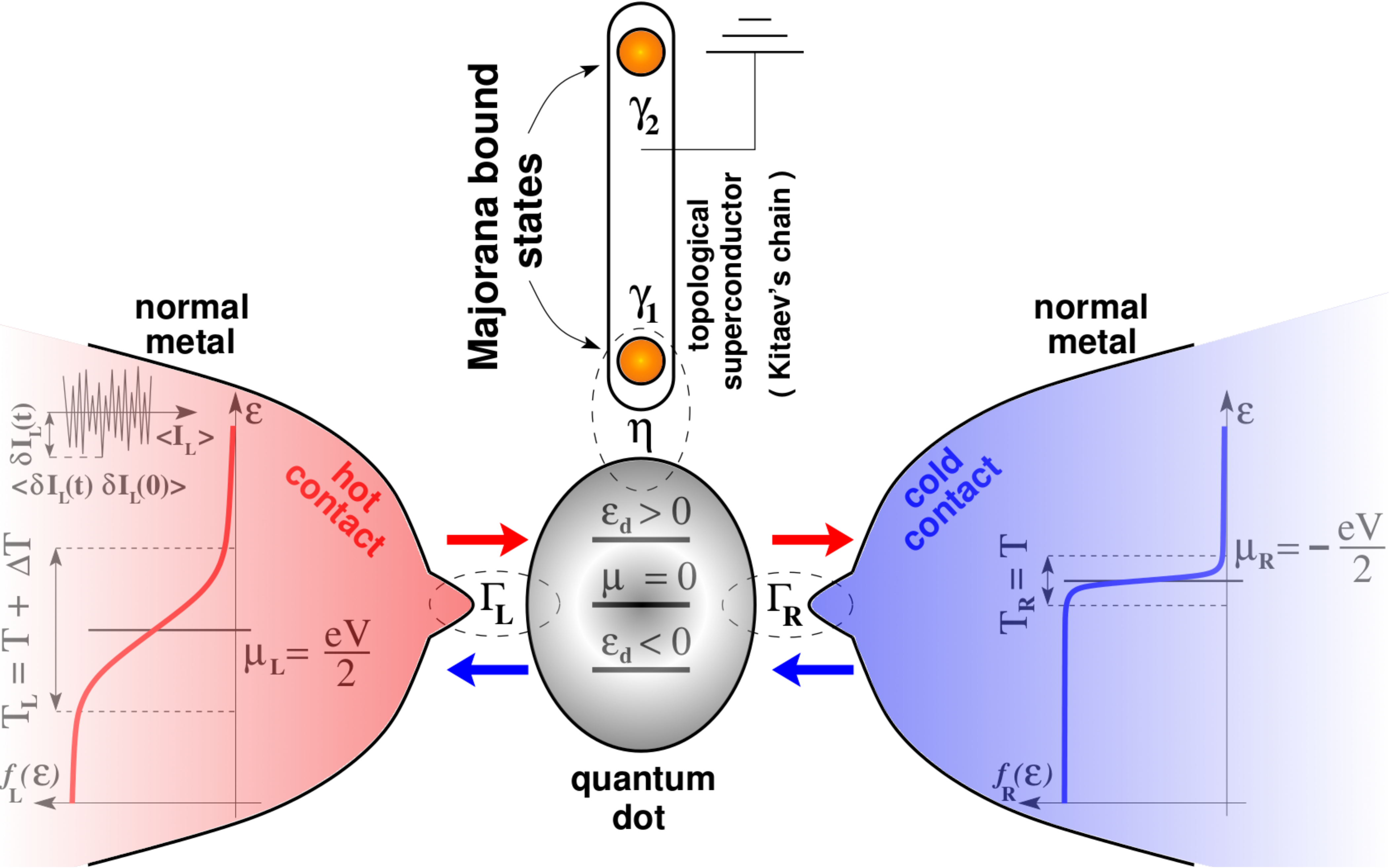}
\caption{\label{figure_1} The physical system represents a quantum dot whose
  single-particle energy level $\epsilon_d$ may be tuned by an external gate
  voltage to make the quantum dot filled ($\epsilon_d<0$) or empty
  ($\epsilon_d>0$). The left ($L$) and right ($R$) contacts are normal metals
  assumed to be in equilibrium characterized by the Fermi-Dirac distributions
  $f_{L,R}(\varepsilon)$ specified by the chemical potentials $\mu_{L,R}$ and
  the temperatures $T_{L,R}$. They are coupled via tunneling interactions
  (with the strengths $\Gamma_{L,R}$) to the quantum dot. An external bias
  voltage $V$ is applied to the contacts, $\mu_L-\mu_R=eV<0$. The left contact
  is hot (shown as red) while the right one is cold (shown as blue),
  $T_L=T+\Delta T$, $T_R=T$, $\Delta T\geqslant 0$. The grounded topological
  superconductor, implemented, {\it e.g}, by means of the Kitaev's chain,
  supports two Majorana modes $\gamma_{1,2}$ at its ends. One of these ends,
  namely the one supporting the Majorana mode $\gamma_1$, is coupled via
  another tunneling interaction (with the strength $\eta$) to the quantum
  dot. The blue and red arrows illustrate schematically the electric current
  flows excited by the external bias voltage $V$ and the temperature
  difference $\Delta T$, respectively. The noise of the electric current
  $S^>(t)$ and the mean electric current $I$ are measured in the hot (left)
  contact, $S^>(t)=\langle \delta I_L(t)\delta I_L(0)\rangle$, $I=\langle I_L\rangle$.} 
\end{figure}

To explore Majorana thermoelectric transport in a nanoscopic setup we focus on
a particular example of a quantum dot system shown in Fig. \ref{figure_1}. It
includes a noninteracting quantum dot with a single-particle energy level
$\epsilon_d$ which is spin nondegenerate. Physical realization may represent a 
quantum dot subjected to a Zeeman field leading to a spin polarized energy
level $\epsilon_d$. Additionally, the position of this energy level with
respect to the chemical potential may be controlled by an external gate
voltage so that the quantum dot is filled when $\epsilon_d<0$ and empty when
$\epsilon_d>0$. The Hamiltonian of the isolated quantum dot is
\begin{equation}
\hat{H}_d=\epsilon_d d^\dagger d,
\label{Ham_QD}
\end{equation}
where $d$ and $d^\dagger$ are, respectively, the quantum dot Dirac fermion
annihilation and creation operators with the usual anticommutation relations,
$\{d,d\}=0$, $\{d,d^\dagger\}=1$. If, as mentioned above, a Zeeman field is
applied to the quantum dot, it will destroy the correlation effects due to the
Kondo effect \cite{Hewson_1997} so that the latter is not important for the
Majorana physics discussed below. Moreover, as is well known, the Kondo effect
in the present setup would require the quantum dot to be filled
($\epsilon_d<0$). However, for the case of the empty quantum dot
($\epsilon_d>0$) the Kondo effect does not arise. On the other side, as will
be shown below (see also Ref. \cite{Smirnov_2017} for pure electric Majorana
noise), the thermoelectric Majorana noise is universal and it does not depend
on whether $\epsilon_d<0$ or $\epsilon_d>0$. Therefore, in order to completely
exclude the Kondo effect from the physical setup, we will below always
consider the case $\epsilon_d>0$.

The setup also contains two contacts labeled as left ($L$) and right
($R$). They represent noninteracting normal metals which are assumed to be in
equilibrium characterized by the Fermi-Dirac distributions:
\begin{equation}
f_{L,R}(\epsilon)=\frac{1}{\exp\bigl(\frac{\epsilon-\mu_{L,R}}{T_{L,R}}\bigl)+1},
\label{cont_FD}
\end{equation}
where $\mu_{L,R}$ and $T_{L,R}$ are, respectively, the chemical potentials and
temperatures of the contacts. An external bias voltage $V$ may be applied to
the contacts so that $\mu_L-\mu_R=eV$, where $e$ is the electron charge, and
we will assume the symmetric bias, $\mu_{L,R}=\pm eV/2$, with $eV<0$. The
temperatures of the contacts are $T_L=T+\Delta T$, $T_R=T$ with
$T\geqslant 0$, $\Delta T\geqslant 0$, {\it i.e.} the left contact is hot and
the right contact is cold. We consider a typical setup where the two contacts
are characterized by the same set of quantum numbers $k$ and energy spectrum
$\epsilon_k$. In this case the Hamiltonian of the contacts is
\begin{equation}
\hat{H}_c=\sum_{l=\{L,R\}}\sum_k\epsilon_k c_{lk}^\dagger c_{lk},
\label{Ham_C}
\end{equation}
where $c_{lk}$ and $c_{lk}^\dagger$ are, respectively, the contacts Dirac
fermion annihilation and creation operators with the usual anticommutation
relations, $\{c_{lk},c_{l'k'}\}=0$,
$\{c_{lk},c_{l'k'}^\dagger\}=\delta_{ll'}\delta_{kk'}$. If the quantum dot
energy level $\epsilon_d$ is spin polarized, {\it i.e.} it has a definite spin
$\sigma$ realized via a Zeeman splitting, as mentioned above, then
Eq. (\ref{Ham_C}) describes the electrons in the normal metals with the same
spin $\sigma$. Additionally, we assume massive contacts so that their energy
spectrum is continuous and their density of states per spin may be
characterized by a constant $\nu_c/2$ within the energy range relevant for
transport.

The quantum dot interacts with the contacts via tunneling,
\begin{equation}
\hat{H}_{d-c}=\sum_{l=\{L,R\}}\sum_k T_{lk} c_{lk}^\dagger d+\text{H.c.},
\label{Ham_T}
\end{equation}
where the dependence of the tunneling matrix elements on the contacts quantum
numbers is usually neglected, $T_{lk}=T_l$. The strength of the tunneling
between the quantum dot and the left or right contact is characterized by the
quantity $\Gamma_l\equiv\pi\nu_c|T_l|^2$ while the total tunneling strength is
defined as $\Gamma\equiv\sum_{l=\{L,R\}}\Gamma_l=\Gamma_L+\Gamma_R$.

The final constituent of the setup is a grounded topological superconductor
implementing the Kitaev's one-dimensional chain supporting two Majorana zero
modes localized at its edges. Due to a finite length of the topological
superconductor these Majorana bound states may have a finite overlap with a
characteristic energy scale $\xi$. For large values of $\xi$ the two Majorana
fermions merge and behave as a single Dirac fermion while for small values of
$\xi$ the low-energy physics is essentially governed by Majorana degrees of
freedom. The effective low-energy Hamiltonian of the topological
superconductor is
\begin{equation}
\hat{H}_{tsc}=\frac{\mathrm{i}}{2}\xi\gamma_2\gamma_1,
\label{Ham_TS}
\end{equation}
where $\gamma_{1,2}$ are the Majorana fermion annihilation operators identical
to the corresponding creation operators,
$\gamma_{1,2}^\dagger=\gamma_{1,2}$. These annihilation operators satisfy the
anticommutation relations $\{\gamma_i,\gamma_j\}=2\delta_{ij}$ implying that
the associative algebra generated by the set $\{1,\gamma_1,\gamma_2\}$ is the
Clifford algebra \cite{Fuchs_1997}.

The quantum dot interacts with the topological superconductor via tunneling
involving only the Majorana mode $\gamma_1$ as has been suggested in numerous
literature (see, {\it e.g.},
Refs. \cite{Cheng_2014,Liu_2015,Liu_2015a,Lopez_2014}),
\begin{equation}
\hat{H}_{d-tsc}=\eta^*d^\dagger\gamma_1+\text{H.c.},
\label{Ham_MT}
\end{equation}
where the absolute value $|\eta|$ of the tunneling matrix element
characterizes the strength of the tunneling between the quantum dot and the
topological superconductor.

To explore nonlinear thermoelectric response of the system specified by the
Hamiltonian
\begin{equation}
\hat{H}=\hat{H}_d+\hat{H}_c+\hat{H}_{d-c}+\hat{H}_{tsc}+\hat{H}_{d-tsc}
\label{Ham}
\end{equation}
it is convenient to resort to the Keldysh field integral framework
\cite{Kamenev_1999,Altland_2010} as a general tool providing various
correlation functions in a simple systematic way. To this end we write down
the source dependent Keldysh generating functional:
\begin{equation}
Z[J_l(t)]=\int\mathcal{D}[\bar{\theta}(t),\theta(t)]e^{\frac{\mathrm{i}}{\hbar}S_K[\bar{\theta}(t),\theta(t);J_l(t)]},
\label{KJF}
\end{equation}
where
$\{\bar{\theta}(t),\theta(t)\}=\{\bar{\psi}(t),\psi(t);\bar{\phi}_{lk}(t),\phi_{lk}(t);\bar{\zeta}(t),\zeta(t)\}$
is the set of the Grassmann fields of, respectively, the quantum dot, contacts
and topological superconductor while $J_l(t)$ is the source field. Note, the
fundamental normalization $Z[J_l(t)=0]=1$. In Eq. (\ref{KJF}) the time
argument $t$ runs over the Keldysh closed time contour $C_K$, $t\in C_K$.

The total Keldysh action $S_K[\bar{\theta}(t),\theta(t);J_l(t)]$ consists of
the Keldysh actions of the isolated quantum dot, contacts and topological
superconductor, the tunneling actions describing the interaction of the
quantum dot with the contacts and topological superconductor as well as the
source action.

The Keldysh actions of the isolated quantum dot, $S_d[\bar{\psi}(t),\psi(t)]$,
contacts, $S_c[\bar{\phi}_{lk}(t),\phi_{lk}(t)]$, and topological
superconductor, $S_{tsc}[\bar{\zeta}(t),\zeta(t)]$, have the conventional form
of upper triangular $2\times 2$ matrices in the retarded-advanced space with
the upper/lower diagonal elements representing the inverse retarded/advanced
Green's functions of the corresponding isolated system while the upper
off-diagonal elements of these matrices have the form
$\mathrm{i}\,\delta\,(1-2f)$ where $\delta\rightarrow 0^+$ and $f$ is
the corresponding Fermi-Dirac distribution which, {\it e.g.}, for the contacts
is given by Eq. (\ref{cont_FD}).

The tunneling actions describing the interactions of the quantum dot with the
contacts and topological superconductor are, respectively, given as
\begin{equation}
\begin{split}
&S_{d-c}[\bar{\psi}(t),\psi(t);\bar{\phi}_{lk}(t),\phi_{lk}(t)]=\\
&=-\int_{-\infty}^\infty dt\sum_{l=\{L,R\}}\sum_{k}\bigl\{T_l[\bar{\phi}_{lk+}(t)\psi_+(t)-\\
&-\bar{\phi}_{lk-}(t)\psi_-(t)]+\text{G.c.}\bigl\},
\end{split}
\label{Tunn_act_dc}
\end{equation}
\begin{equation}
\begin{split}
&S_{d-tsc}[\bar{\psi}(t),\psi(t);\bar{\zeta}(t),\zeta(t)]=\\
&=-\int_{-\infty}^\infty dt\bigl\{\eta^*[\bar{\psi}_+(t)\zeta_+(t)+\bar{\psi}_+(t)\bar{\zeta}_+(t)-\\
&-\bar{\psi}_-(t)\zeta_-(t)-\bar{\psi}_-(t)\bar{\zeta}_-(t)]+\text{G.c.}\bigl\},
\end{split}
\label{Tunn_act_ds}
\end{equation}
where in the right hand side the time argument $t$ runs over the real axis and
the subindex $+/-$ denotes the forward/backward branches of the Keldysh closed
time contour. The abbreviation $\text{G.c.}$ stands for the Grassmann
conjugation, the generalization of the Hermitian conjugation for the case when
Grassmann variables are involved in a mathematical expression.

Finally, the source action is
\begin{equation}
\begin{split}
&S_{scr}[\bar{\psi}(t),\psi(t);\bar{\phi}_{lk}(t),\phi_{lk}(t);J_l(t)]=\\
&=-\int_{-\infty}^\infty dt\sum_{l=\{L,R\}}\sum_{q=\{+,-\}}J_{lq}(t)I_{lq}(t),
\end{split}
\label{Scr_act}
\end{equation}
where $I_{lq}(t)$ is the field representing the electric current in the left
($l=L$) or right ($l=R$) contact on the forward ($q=+$) or backward ($q=-$)
branch of the Keldysh closed time contour:
\begin{equation}
I_{lq}(t)\equiv\frac{\mathrm{i}e}{\hbar}\sum_k[T_l\bar{\phi}_{lkq}(t)\psi_q(t)-\text{G.c.}].
\label{Curr_fld}
\end{equation}

Using the total Keldysh action,
\begin{equation}
\begin{split}
&S_K[\bar{\theta}(t),\theta(t);J_l(t)]=\\
&=S_d[\bar{\psi}(t),\psi(t)]+S_c[\bar{\phi}_{lk}(t),\phi_{lk}(t)]+\\
&+S_{tsc}[\bar{\zeta}(t),\zeta(t)]+S_{d-c}[\bar{\psi}(t),\psi(t);\bar{\phi}_{lk}(t),\phi_{lk}(t)]+\\
&+S_{d-tsc}[\bar{\psi}(t),\psi(t);\bar{\zeta}(t),\zeta(t)]+\\
&+S_{scr}[\bar{\psi}(t),\psi(t);\bar{\phi}_{lk}(t),\phi_{lk}(t);J_l(t)],
\end{split}
\label{Tot_Keld_act}
\end{equation}
one can easily obtain the mean electric current in the left ($l=L$) or
right ($l=R$) contact by taking the first derivative of $Z[J_l(t)]$ with
respect to the corresponding source field:
\begin{equation}
\langle I_l\rangle\equiv\langle I_{lq}(t)\rangle_{S_K}=\mathrm{i}\hbar\frac{\delta Z[J_l(t)]}{\delta J_{lq}(t)}\biggl|_{J_{lq}(t)=0},
\label{Mean_curr}
\end{equation}
where the angular brackets
$\langle\cdots\rangle_{S_K}$ denote the average of a functional of Grassmann
fields with respect to the total Keldysh action (\ref{Tot_Keld_act}) taken at
zero source field,
\begin{equation}
\begin{split}
&\langle\mathcal{F}[\bar{\theta}(t),\theta(t)]\rangle_{S_K}\equiv\\
&\equiv\int\mathcal{D}[\bar{\theta}(\tilde{t}),\theta(\tilde{t})]e^{\frac{\mathrm{i}}{\hbar}S_K[\bar{\theta}(\tilde{t}),\theta(\tilde{t})]}\mathcal{F}[\bar{\theta}(t),\theta(t)],
\end{split}
\label{Aver_zero_scr_fld}
\end{equation}
\begin{equation}
\begin{split}
&S_K[\bar{\theta}(t),\theta(t)]\equiv S_K[\bar{\theta}(t),\theta(t);J_l(t)=0]=\\
&=S_d[\bar{\psi}(t),\psi(t)]+S_c[\bar{\phi}_{lk}(t),\phi_{lk}(t)]+\\
&+S_{tsc}[\bar{\zeta}(t),\zeta(t)]+S_{d-c}[\bar{\psi}(t),\psi(t);\bar{\phi}_{lk}(t),\phi_{lk}(t)]+\\
&+S_{d-tsc}[\bar{\psi}(t),\psi(t);\bar{\zeta}(t),\zeta(t)].
\end{split}
\label{Tot_Keld_act_zero_scr_fld}
\end{equation}

The values of $q$ and $t$ in Eq. (\ref{Mean_curr}) are arbitrary since the
final result does not depend on the choice of a branch of the Keldysh closed
time contour as well as on the choice of an instant of time on that branch
because we consider only stationary nonequilibrium states. It is
straightforward to verify that Eq. (\ref{Mean_curr}) leads to the
Meir-Wingreen result \cite{Meir_1992}.

In a similar way one can obtain the current-current correlator by taking the
second derivative of $Z[J_l(t)]$ with respect to the corresponding source
fields:
\begin{equation}
\begin{split}
&\langle I_l(t)I_{l'}(t')\rangle\equiv\langle I_{l-}(t)I_{l'+}(t')\rangle_{S_K}=\\
&=(\mathrm{i}\hbar)^2\frac{\delta^2Z[J_l(t)]}{\delta J_{l-}(t)\delta J_{l'+}(t')}\biggl|_{J_{lq}(t)=0}.
\end{split}
\label{Curr_curr_corr}
\end{equation}
The technical steps one makes in the calculation of the second derivative in
Eq. (\ref{Curr_curr_corr}) are mainly straightforward. One should only take
proper care of the fact that due to the presence of the topological
superconductor there will appear anomalous contributions when averaging
products of four Grassmann fields of the quantum dot. To facilitate one's
derivation of the anomalous terms and to avoid technicalities in the main text
we provide the main details relevant for the calculation of the second
derivative in Eq. (\ref{Curr_curr_corr}) in the appendix.

The first and second derivatives of the Keldysh generating functional are
enough to explore nonlinear Majorana thermoelectric response of the system in
terms of the mean value and fluctuations of the electric current.
\section{Universal Majorana thermoelectric transport}\label{un_ma_thrmel_tr}
For definiteness below we focus on transport measurements in the left ($L$)
contact which is assumed to be hot. The temperature difference $\Delta T$
between the contacts may be parameterized by a thermal voltage $V_T$,
\begin{equation}
V_T\equiv\frac{k_\text{B}\Delta T}{e},
\label{V_T}
\end{equation}
where $k_\text{B}$ is the Boltzmann constant.

Additionally, we will consider the situation when the quantum dot couples
symmetrically to the left and right contacts, $\Gamma_L=\Gamma_R=\Gamma/2$.

To solely focus on universal Majorana physics we explore the regime dominated
by the Majorana tunneling,
\begin{equation}
|\eta|>\text{max}\{|\epsilon_d|,|eV|,eV_T,k_\text{B}T,\Gamma,\xi\}.
\label{M_tunn_reg}
\end{equation}
Here $\epsilon_d$ may be tuned by an external gate voltage to satisfy
(\ref{M_tunn_reg}). Similarly, the bias voltage $V$, thermal voltage $V_T$ and
temperature $T$ may be externally adjusted to satisfy (\ref{M_tunn_reg}). The
Majorana overlap $\xi$ will be small if the topological superconductor in the
setup is chosen to be long enough so that the two Majorana fermions do not
merge into a single Dirac fermion during the tunneling into the quantum
dot. In modern experiments \cite{Goldhaber-Gordon_1998} the values of $|\eta|$
and $\Gamma$ are readily controlled via external gates whose potentials can
increase or decrease the height of the potential barriers between the quantum
dot and contacts in order to vary $\Gamma$ as well as between the quantum dot
and topological superconductor in order to vary $|\eta|$. In this way in
modern laboratories $|\eta|$ may be increased while $\Gamma$ may be decreased
in order to reach the condition in (\ref{M_tunn_reg}). Thus, besides
its theoretical importance, the Majorana transport regime, specified by
(\ref{M_tunn_reg}), is of particular experimental interest.

Before we start to discuss our results it is important to mention how we
obtain various nonlinear response coefficients. In what follows we use the
formalism presented in Section \ref{fld_thr} and the appendix and perform
numerical calculations of corresponding integrals in the energy domain to
obtain the mean electric current and current-current correlations. This
formalism allows one to compute not only the mean electric current and
current-current correlations but also various derivatives of these
quantities. The calculations of the derivatives may be performed in two
different ways. On one side, one can calculate derivatives using finite
differences. In this case one computes the mean electric current and
current-current correlations on a fine grid of the bias voltage and thermal
voltage and after that applies conventional finite difference schemes for the
first and second derivatives. On the other side, one notices (see the
appendix) that the dependence of both the mean electric current and
current-current correlations on the bias voltage and thermal
voltage enters through the Keldysh components of the Green's functions, more
specifically, through the Fermi-Dirac distributions
(\ref{cont_FD}). Therefore, one can first calculate various derivatives of
corresponding integrands. This simply reduces to analytical differentiations
of the Fermi-Dirac distributions (\ref{cont_FD}). After that one performs
numerical integration of these differentiated integrands which now involve
various analytical derivatives of the Fermi-Dirac distributions
(\ref{cont_FD}). We find that in all of our calculations these two ways of
computing derivatives give the same results within a good numerical
precision. This is a very good test that our numerical results are
reliable. However, in order to reach higher precision to obtain analytical
expressions, which will be discussed below, we prefer to use the second way
involving analytical differentiations of the Fermi-Dirac distributions
(\ref{cont_FD}).
\begin{figure}
\includegraphics[width=8.0 cm]{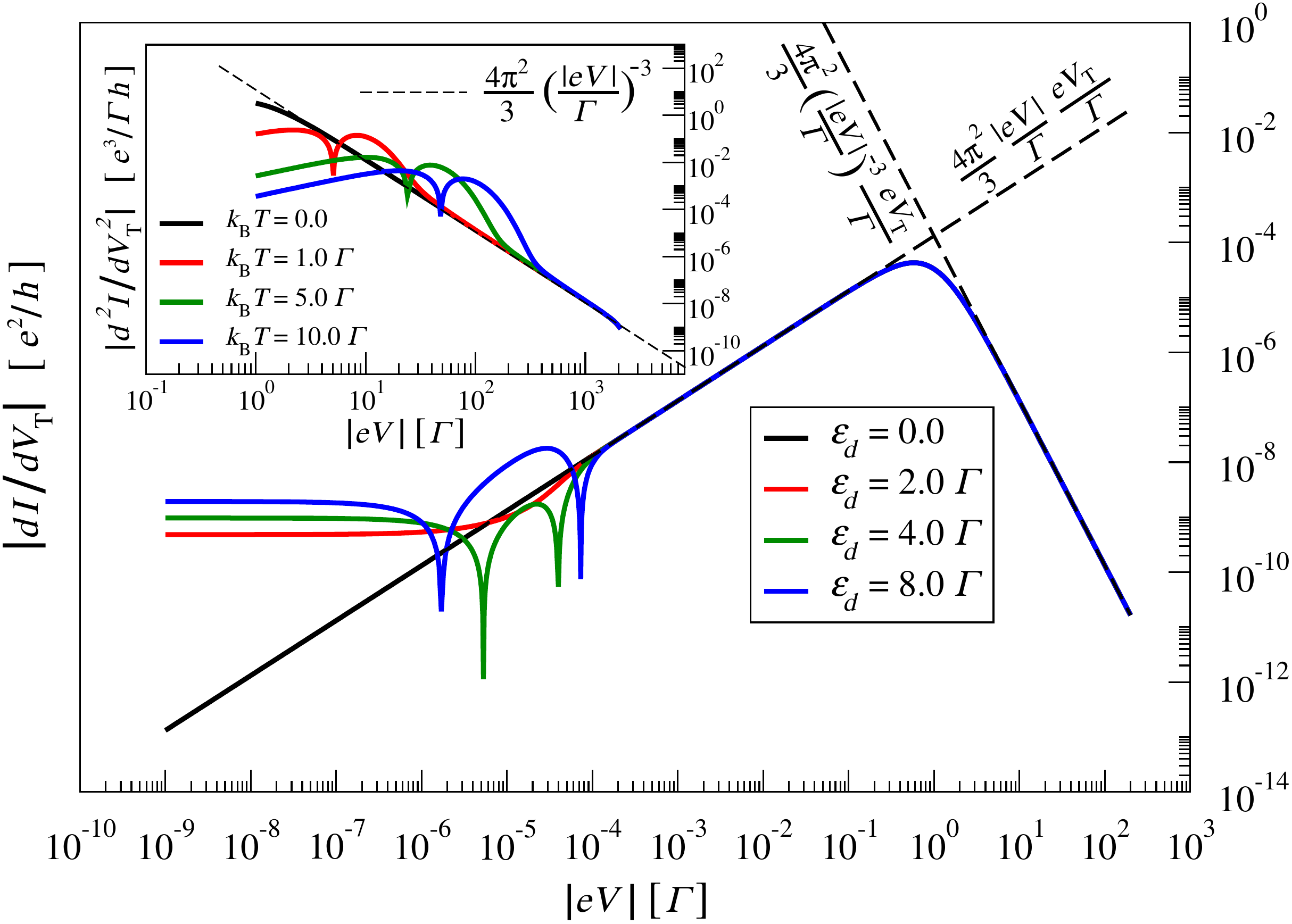}
\caption{\label{figure_2} The magnitude of the first derivative of the mean
  electric current with respect to the thermal voltage,
  $|\partial I(V,V_T)/\partial V_T|$, as a function of the bias voltage $V$ at
  $T=0$, $eV_T/\Gamma=10^{-5}$ and for different gate voltages, specified by
  the quantum dot single-particle energy level $\epsilon_d$. The Majorana
  fermions overlap weakly, $\xi/\Gamma=10^{-4}$. The strength of the Majorana
  tunneling is $|\eta|/\Gamma=10^3$. Here the dips in the non-universal
  ($|eV|\lesssim\xi$) behavior of the curves with $\epsilon_d/\Gamma=4.0$ and
  $\epsilon_d/\Gamma=8.0$ correspond to those values of the bias voltage where
  $\partial I(V,V_T)/\partial V_T=0$, that is to those values of $V$ where
  $\partial I(V,V_T)/\partial V_T$ changes its sign. The inset shows the
  robustness of the universal (independent of $\epsilon_d$) behavior of the
  second derivative of the mean electric current with respect to the thermal
  voltage at large bias voltages when the temperature increases. At higher
  temperatures there are values of the bias voltage where
  $\partial^2I(V,V_T)/\partial V_T^2=0$, that is points where
  $\partial^2I(V,V_T)/\partial V_T^2$ changes its sign. Since we use the
  logarithmic scale, we plot the magnitude of the second derivative,
  $|\partial^2I(V,V_T)/\partial V_T^2|$, which displays dips at its zeroes.}
\end{figure}

Another important aspect concerns the analytical asymptotic limits presented
below. They are obtained by inspection of our numerical results. We will use
the term asymptotic limit for an analytical expression if our numerical
results reproduce this analytical expression with any desired numerical
precision by adjusting the physical parameters of the system to satisfy the
inequalities, specifying the regime of applicability of this analytical
expression, with any desired degree of accuracy. In more mathematical terms,
referring to an analytical expression as an asymptotic limit means that the
stronger the inequalities, specifying the regime of applicability of this
analytical expression, are fulfilled the more digits after the decimal point
in our numerical results reproduce this analytical expression. In other words,
for the theoretical model presented in Section \ref{fld_thr} the asymptotic
limits presented below are analytical expressions to which numerical results
converge when one gradually increases the numerical precision and no further
approximation to the theoretical model in Section \ref{fld_thr} is assumed.

Finally, we would like to specify what we understand under universality of our
results which will be presented below. We will call a quantity universal if
this quantity is independent of the parameters characterizing exclusively the
quantum dot. In our case such a parameter is $\epsilon_d$ which is tuned by an
external gate voltage. This is experimentally relevant because in a realistic
experiment one may easily vary the gate voltage and observe which transport
quantities do not change in response to this variation of the gate
voltage. Similar universality happens in the Kondo effect \cite{Hewson_1997}
where universality is understood as independence of response coefficients on
$\epsilon_d$ which enters only through the Kondo temperature $T_K$. The
difference between the Majorana universality and Kondo universality is in the
scaling. For the present case of the Majorana universality the scaling is
given by $\Gamma$ whereas for the Kondo universality it is given by the Kondo
temperature $T_K$ which is also a function of $\Gamma$. In all of our
calculations presented below we use $\epsilon_d>0$. More specifically, to
perform concrete calculations we put $\epsilon_d=8\Gamma$ but universal
results do not change if one uses other values for $\epsilon_d$ satisfying the
condition in (\ref{M_tunn_reg}). Using positive values for $\epsilon_d$ one
puts the quantum dot in the empty orbital regime \cite{Hewson_1997}. As a
result, in a realistic experiment the Kondo effect is switched off and one
observes only the Majorana universality. Note, that in the absence of the
Majorana bound states one would observe strong dependence of response
coefficients on $\epsilon_d$ for bias voltages
$|eV|\sim\epsilon_d$. Therefore, observing independence of transport
quantities on the gate voltage for any bias voltage already provides
information about signatures of unpaired Majorana fermions. Of course, such a
quantity, which is universal in the above sense, may depend on the tunneling
coupling to the contacts, $\Gamma$, or on the tunneling coupling to the
topological superconductor, $\eta$, or on both. Thus the universality
understood in the above sense is only a qualitative Majorana signature but not
quantitative. For example, a response coefficient may have a universal
asymptotic behavior at high bias voltages. Let us assume that this universal
asymptotic behavior is given by a certain universal (independent of
$\epsilon_d$) function of $eV/\Gamma$. Although the universality of this
function is a qualitative signature of unpaired Majorana fermions, the values
of the coefficients in the expansion of this universal function in powers of
$eV/\Gamma$ will depend on the definition of $\Gamma$ and cannot serve as a
quantitative signature of unpaired Majorana fermions. To provide quantitative
signatures of unpaired Majorana fermions we define truly universal
quantities. We will call a quantity truly universal if this quantity does not
depend on both the parameters of the quantum dot, such as $\epsilon_d$, and
the tunneling couplings to the contacts, $\Gamma$, as well as to the
topological superconductor, $\eta$. For example, if there are two universal
quantities, the ratio between these two quantities may be truly universal
because $\Gamma$ and $\eta$ may cancel out in this ratio which therefore
becomes a unique quantitative signature of unpaired Majorana fermions. Below
in the discussion of our results we will always distinguish between universal
and truly universal results.
\subsection{Nonequilibrium response of the mean thermoelectric current}\label{res_curr}
Let us first examine the universal Majorana signatures one can detect in the
thermoelectric response of the mean electric current $I(V,V_T)$ as a function
of the bias voltage and thermal voltage. It is obtained from
Eq. (\ref{Mean_curr}) with $l=L$, that is $I(V,V_T)\equiv\langle I_L\rangle$.
\begin{figure}
\includegraphics[width=8.0 cm]{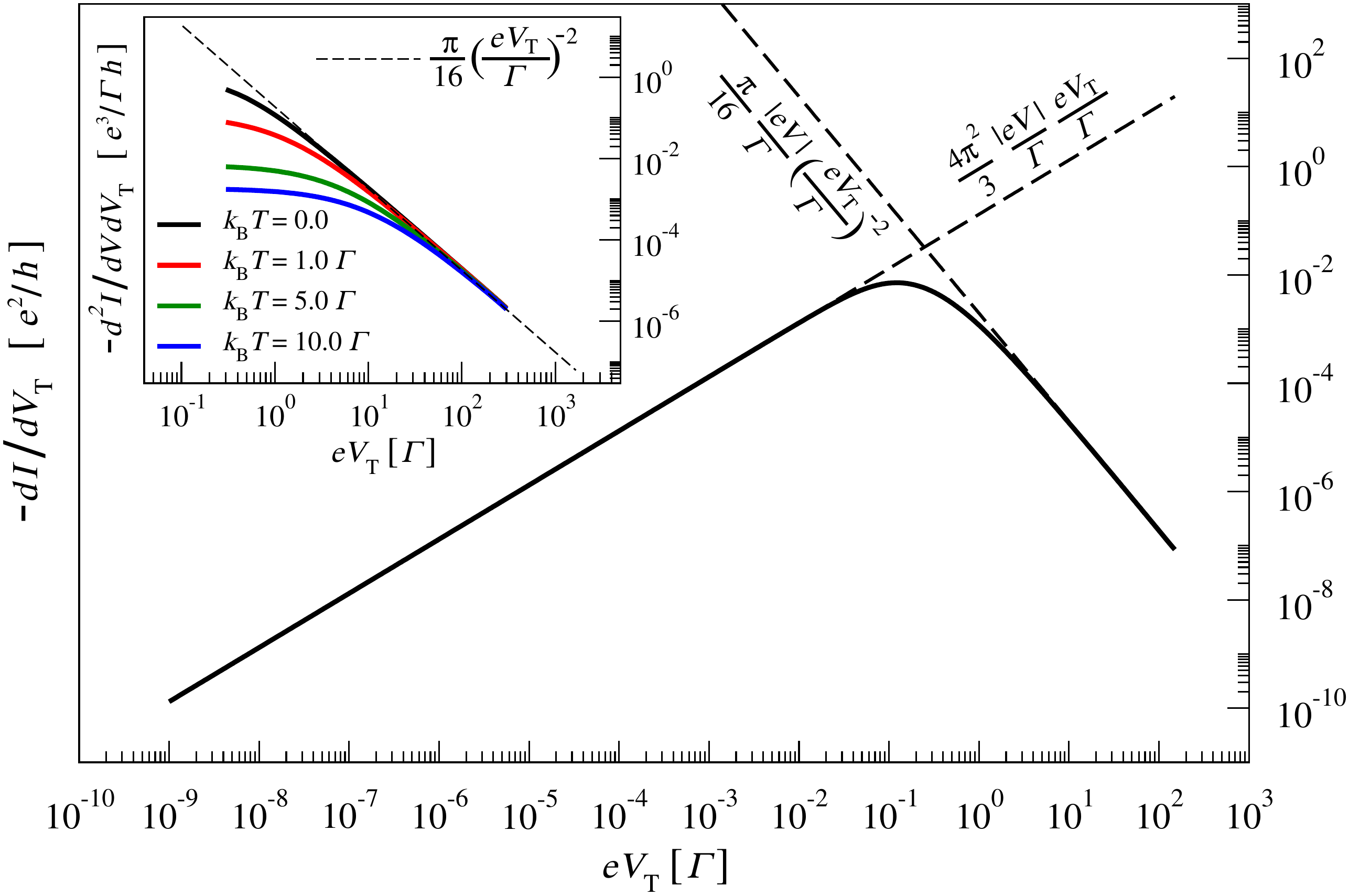}
\caption{\label{figure_3} The first derivative of the mean current with
  respect to the thermal voltage, $\partial I(V,V_T)/\partial V_T$, as a
  universal (independent of $\epsilon_d$) function of the thermal voltage
  $V_T$ at $T=0$, $|eV|/\Gamma=10^{-2}$. The other parameters have the same
  values as in Fig. \ref{figure_2}. The inset shows the robustness of the
  universal (independent of $\epsilon_d$) behavior of the second mixed
  derivative $\partial^2I(V,V_T)/\partial V\partial V_T$ at large thermal
  voltages when the temperature increases.}
\end{figure}

In Fig. \ref{figure_2} we show the magnitude of the thermoelectric coefficient
representing the first derivative of the mean electric current with respect to
the thermal voltage, $|\partial I(V,V_T)/\partial V_T|$, as a function of the
bias voltage $V$ at zero temperature for the case $eV_T\ll\Gamma$ and for
different gate voltages parameterized by the values of $\epsilon_d$. One can
see that despite the fact that the Majorana fermions are well separated
($\xi/\Gamma=10^{-4}$) and the Majorana tunneling is strong
($|\eta|/\Gamma=10^3$), the curves do not collapse on a single universal
(independent of $\epsilon_d$) curve. We find that only for $|eV|\gg\xi$ the
curves do collapse on a single universal curve whose shape does not depend on
the gate voltage which regulates the value of $\epsilon_d$. As mentioned in
Section \ref{fld_thr}, we show only the case $\epsilon_d>0$ since universal
Majorana signatures do not depend on both the magnitude and sign of
$\epsilon_d$. At the same time for $\epsilon_d>0$ the quantum dot is in the
empty orbital regime which excludes the Kondo resonance \cite{Hewson_1997}
from the physical setup. Thus we conclude that the linear thermoelectric
response, {\it i.e.} response at small bias voltages ($|eV|\ll\xi$), of the
mean electric current provides only non-universal Majorana signatures and,
therefore, it does not represent any interest as a unique universal signature
of Majorana fermions.

To access the universal Majorana physics in the mean
thermoelectric quantities one has to resort to the nonlinear ($|eV|>\xi$)
thermoelectric response. In this universal regime we find the following
asymptotic limits for the thermoelectric coefficient
$\partial I(V,V_T)/\partial V_T$: 
\begin{equation}
\frac{\partial I(V,V_T)}{\partial V_T}=-\frac{e^2}{h}\frac{4\pi^2}{3}\frac{|eV|}{\Gamma}\frac{eV_T}{\Gamma}
\label{It_1}
\end{equation}
for
\begin{equation}
|\eta|\gg\Gamma\gg (|eV|,eV_T),\quad |eV|\gg\xi
\label{It_1_cond}
\end{equation}
and
\begin{equation}
\frac{\partial I(V,V_T)}{\partial V_T}=-\frac{e^2}{h}\frac{4\pi^2}{3}\biggl(\frac{|eV|}{\Gamma}\biggl)^{-3}\frac{eV_T}{\Gamma}
\label{It_2}
\end{equation}
for
\begin{equation}
|\eta|\gg |eV|\gg\Gamma\gg eV_T,\quad\Gamma\gg\xi.
\label{It_2_cond}
\end{equation}

The universal laws given by Eqs. (\ref{It_1}) and (\ref{It_2}) are depicted by
dashed lines in Fig. \ref{figure_2}.

The inset in Fig. \ref{figure_2} shows the universal (independent of
$\epsilon_d$) Majorana behavior of the thermoelectric coefficient representing
the second derivative of the mean electric current with respect to the thermal
voltage, $\partial^2 I(V,V_T)/\partial V_T^2$, as a function of the bias
voltage $V$ in the high-energy regime specified by the inequality in
(\ref{It_2_cond}). This universal Majorana high-energy behavior is independent
of the thermal voltage $V_T$. As one can see, the second derivative
$\partial^2I(V,V_T)/\partial V_T^2$ is very robust with respect to high
temperatures. At high bias voltages it is given by the asymptotic limit,
\begin{equation}
\frac{\partial^2 I(V,V_T)}{\partial V_T^2}=-\frac{e^3}{\Gamma h}\frac{4\pi^2}{3}\biggl(\frac{|eV|}{\Gamma}\biggl)^{-3},
\label{I_tt_2}
\end{equation}
shown as the dashed line in the inset in Fig. \ref{figure_2}. The expression
in Eq. (\ref{I_tt_2}) follows from Eq. (\ref{It_2}) if the inequality in
(\ref{It_2_cond}) is fulfilled. The universal behavior of
$\partial^2I(V,V_T)/\partial V_T^2$ given by Eq. (\ref{I_tt_2}) is almost
unchanged for
$|eV|\gg k_\text{B}T$ when the temperature increases up to
$k_\text{B}T\sim 10\,\Gamma$, or in units of $|\eta|$,
$k_\text{B}T\sim 0.01\,|\eta|$.

As it follows from the above analysis, to focus on universal thermoelectric
Majorana signatures in the mean electric current one has to apply bias
voltages $|eV|\gg\xi$. In Fig. \ref{figure_3} we show the universal Majorana
behavior of the thermoelectric coefficient representing the first derivative
of the mean electric current with respect to the thermal voltage,
$\partial I(V,V_T)/\partial V_T$, as a function of the thermal voltage $V_T$
at zero temperature and at a fixed value of the bias voltage
$\Gamma\gg |eV|\gg\xi$. For the particular example shown in Fig. \ref{figure_3}
we put $|eV|/\xi=100$, $|eV|/\Gamma=0.01$. As a result, the curve in
Fig. \ref{figure_3} is universal in the whole range of the thermal
voltage. The universal behavior of $\partial I(V,V_T)/\partial V_T$ as a
function of the thermal voltage $V_T$ in the regime specified by the
inequality (\ref{It_1_cond}) is given by the asymptotic limit,
Eq. (\ref{It_1}), shown in Fig. \ref{figure_3} as the dashed line with a
positive slope. However, for
\begin{equation}
|\eta|\gg eV_T\gg\Gamma\gg |eV|\gg\xi
\label{It_3_cond}
\end{equation}
we find the following universal asymptotic limit:
\begin{equation}
\frac{\partial I(V,V_T)}{\partial V_T}=-\frac{e^2}{h}\frac{\pi}{16}\frac{|eV|}{\Gamma}\biggl(\frac{eV_T}{\Gamma}\biggl)^{-2}
\label{It_3}
\end{equation}
which is shown in Fig. \ref{figure_3} as the dashed line with a negative
slope.
\begin{figure}
\includegraphics[width=8.0 cm]{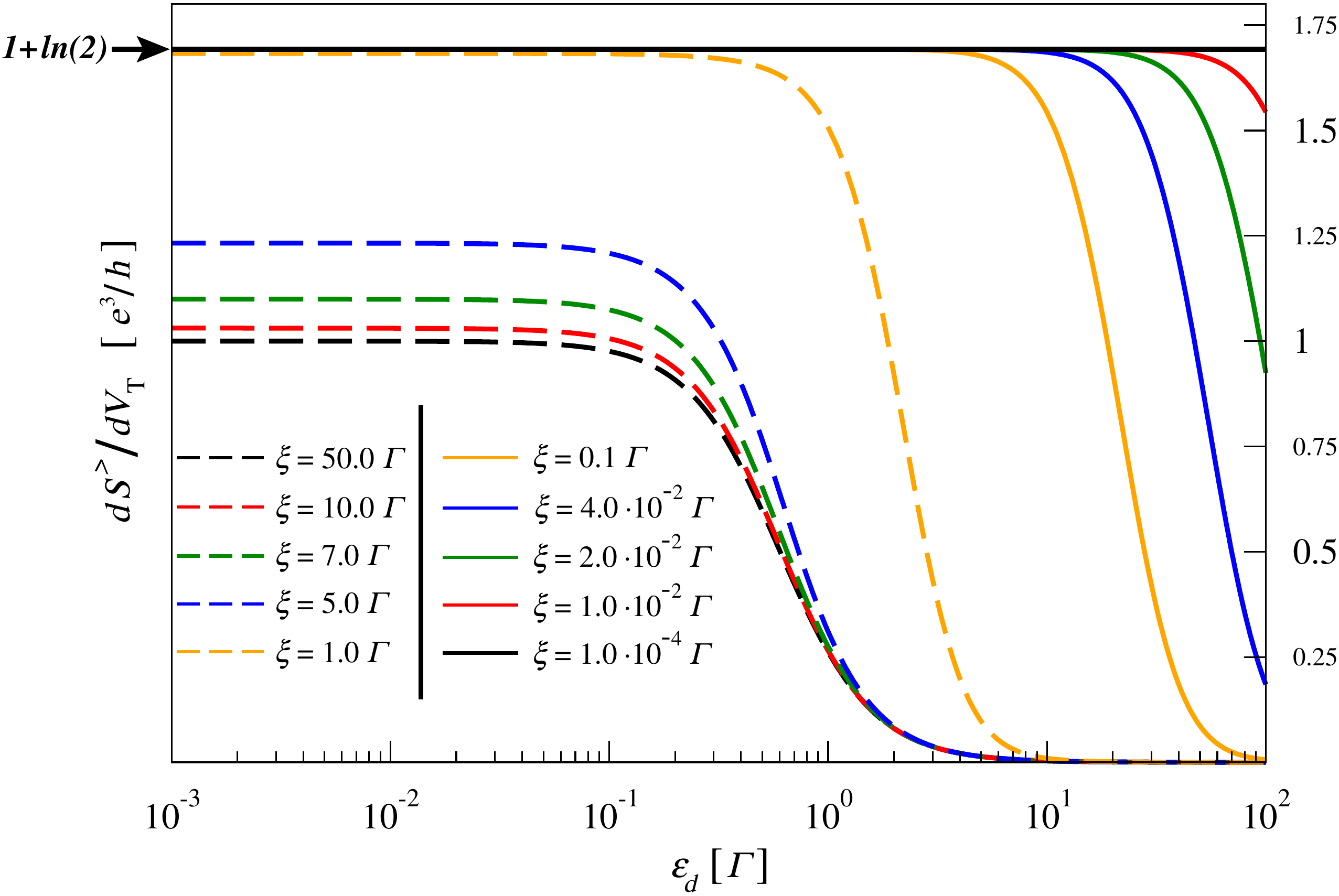}
\caption{\label{figure_4} The first derivative of the greater current-current
  correlator with respect to the thermal voltage,
  $\partial S^>(V,V_T)/\partial V_T$, as a function of the gate voltage,
  specified by the quantum dot single-particle energy level $\epsilon_d$, for
  different values of the Majorana overlap energy $\xi$. Here the parameters
  are as follows $T=0$, $V=0$, $eV_T/\Gamma=10^{-6}$ and $\eta/\Gamma=10^3$.}
\end{figure}

The inset in Fig. \ref{figure_3} shows the universal Majorana behavior of the
thermoelectric coefficient representing the second mixed derivative of the
mean electric current with respect to the thermal voltage and the bias voltage
$\partial^2 I(V,V_T)/\partial V\partial V_T$, as a function of the thermal
voltage $V_T$ in the high-energy regime specified by the inequality in
(\ref{It_3_cond}). The universal Majorana high-energy behavior of
$\partial^2I(V,V_T)/\partial V\partial V_T$ is independent of the bias voltage
$V$. Like the second derivative $\partial^2I(V,V_T)/\partial V_T^2$ as a
function of the bias voltage $V$, the second mixed derivative
$\partial^2I(V,V_T)/\partial V\partial V_T$ as a function of the thermal
voltage $V_T$ is robust with respect to high temperatures. At high thermal
voltages it is given by the asymptotic limit obtained from Eq. (\ref{It_3}) if
the inequality in (\ref{It_3_cond}) is fulfilled. Its universal behavior is
shown by the dashed line in the inset in Fig. \ref{figure_3}. It is protected
by high thermal voltages $eV_T\gg k_\text{B}T$ from thermal destruction
effects as shown in the inset for high temperatures, $k_\text{B}T\sim
10\,\Gamma$ ($k_\text{B}T\sim 0.01\,|\eta|$).
\subsection{Thermoelectric fluctuations beyond linear response}\label{res_noise}
Now we address the universal Majorana signatures present in the thermoelectric
response of the fluctuations of the electric current. To this end we use
\begin{figure}
\includegraphics[width=8.0 cm]{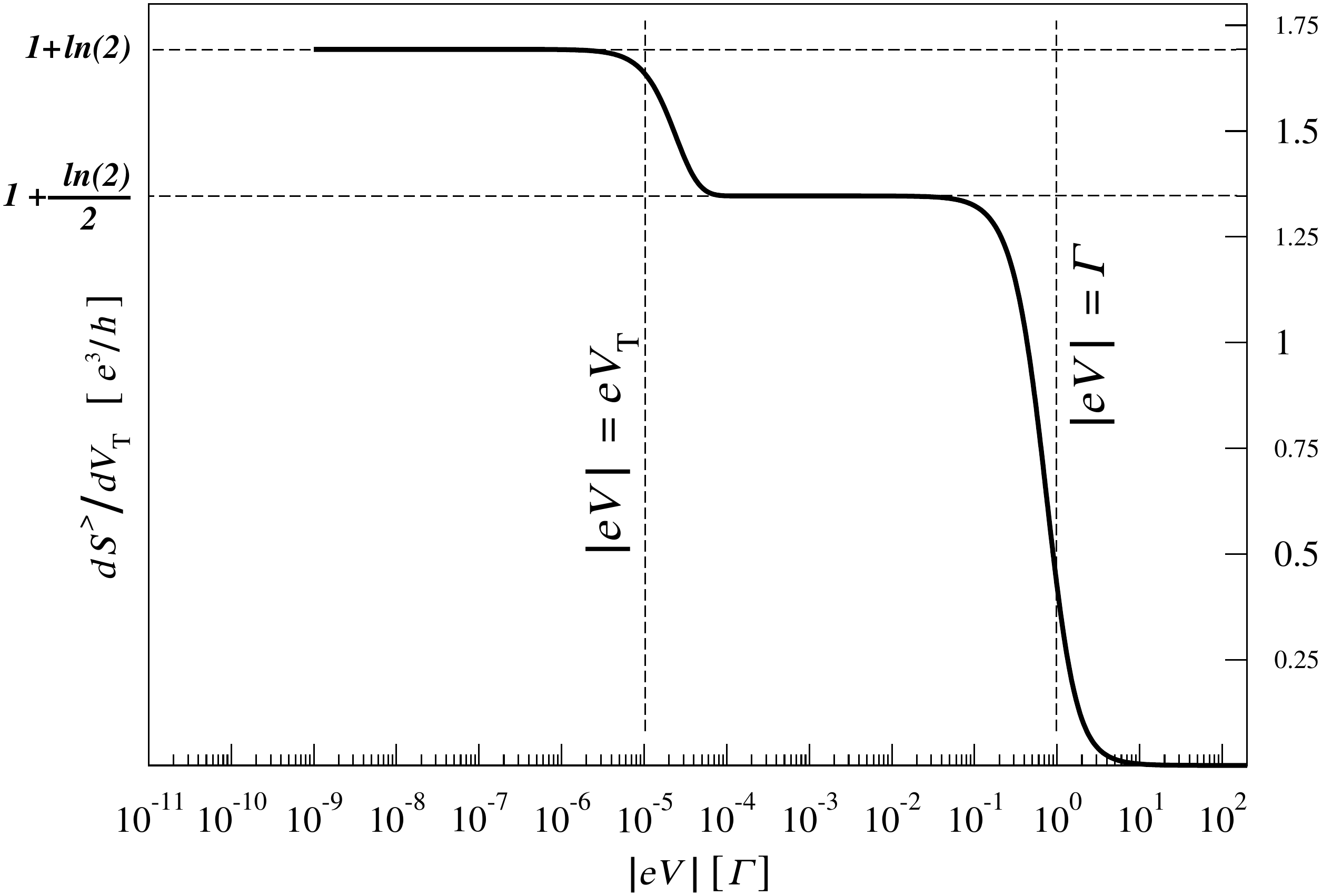}
\caption{\label{figure_5} The first derivative of the greater current-current
  correlator with respect to the thermal voltage,
  $\partial S^>(V,V_T)/\partial V_T$, as a universal (independent of
  $\epsilon_d$) function of the bias voltage $V$ at $T=0$,
  $eV_T/\Gamma=10^{-5}$. The Majorana overlap energy and the strength of the
  Majorana tunneling are $\xi/\Gamma=10^{-4}$ and $|\eta|/\Gamma=10^3$,
  respectively.}
\end{figure}
Eq. (\ref{Curr_curr_corr}) with $l=l'=L$ to obtain the current-current
correlator $\langle I_L(t)I_L(t')\rangle$. Finite deviations of the electric
current from its mean value, $\delta I_L(t)=I_L(t)-I(V,V_T)$, may be
characterized via the greater current-current correlator defined as
$S^>(t,t';V,V_T)\equiv\langle\delta I_L(t)\delta I_L(t')\rangle$. It can be
expressed through the correlator $\langle I_L(t)I_L(t')\rangle$ as follows:
\begin{equation}
S^>(t,t';V,V_T)=\langle I_L(t)I_L(t')\rangle-I^2(V,V_T).
\label{S_grt}
\end{equation}
Due to the stationary nonequilibrium $S^>(t,t';V,V_T)=S^>(t-t';V,V_T)$ and the
physical quantity measured experimentally is the Fourier transform
\begin{equation}
S^>(\omega;V,V_T)=\int_{-\infty}^\infty dt\,e^{\mathrm{i}\omega t}S^>(t;V,V_T).
\label{S_grt_fr}
\end{equation}
Below we focus on the zero frequency noise as a function of the bias voltage
and thermal voltage,
\begin{equation}
S^>(V,V_T)\equiv S^>(\omega=0;V,V_T).
\label{S_gr_zr_fr}
\end{equation}

As we know from the previous subsection the mean Majorana thermoelectric
response becomes universal (independent of $\epsilon_d$) only if the bias
voltage is large enough, $|eV|\gg\xi$. The first question is thus whether the
same happens with the Majorana thermoelectric response of the fluctuations of
the electric current. We find that, in contrast to the mean Majorana
thermoelectric response, the fluctuations of the electric current are
universal in the whole range of the bias voltage. This means that the Majorana
thermoelectric response of the electric current noise is independent of
$\epsilon_d$ at any bias voltage $V$. In Fig. \ref{figure_4} we show the
thermoelectric coefficient representing the first derivative of the greater
noise with respect to the thermal voltage, $\partial S^>(V,V_T)/\partial V_T$,
as a function of the gate voltage, parameterized by the value of $\epsilon_d$,
at zero temperature, zero bias voltage, $eV_T\ll\Gamma$ and different values
of the Majorana overlap energy $\xi$ in the regime specified in
(\ref{M_tunn_reg}) where the Majorana tunneling dominates. As can be seen, for
large values of the Majorana overlap energy there is a very strong dependence
on the gate voltage with the unit plateau at small gate voltages. When $\xi$
decreases the value of the plateau grows and the plateau becomes
wider. Finally, when $\xi$ is very small (the same as in the previous
subsection on the mean electric current, $\xi/\Gamma=10^{-4}$) the Majorana
fermions overlap very weakly. In this situation the value of the plateau
saturates at the asymptotic limit $(e^3/h)[1+\ln(2)]$ and the width of the
plateau gets extremely wide. Therefore, when the Majorana tunneling is strong
enough, so that (\ref{M_tunn_reg}) is satisfied, the fluctuation
thermoelectric coefficient $\partial S^>(V,V_T)/\partial V_T$ becomes
independent of $\epsilon_d$, $\Gamma$ and $\eta$, that is it becomes truly
universal with the value $(e^3/h)[1+\ln(2)]$ (the black horizontal line in
Fig. \ref{figure_4}).
\begin{figure}
\includegraphics[width=8.0 cm]{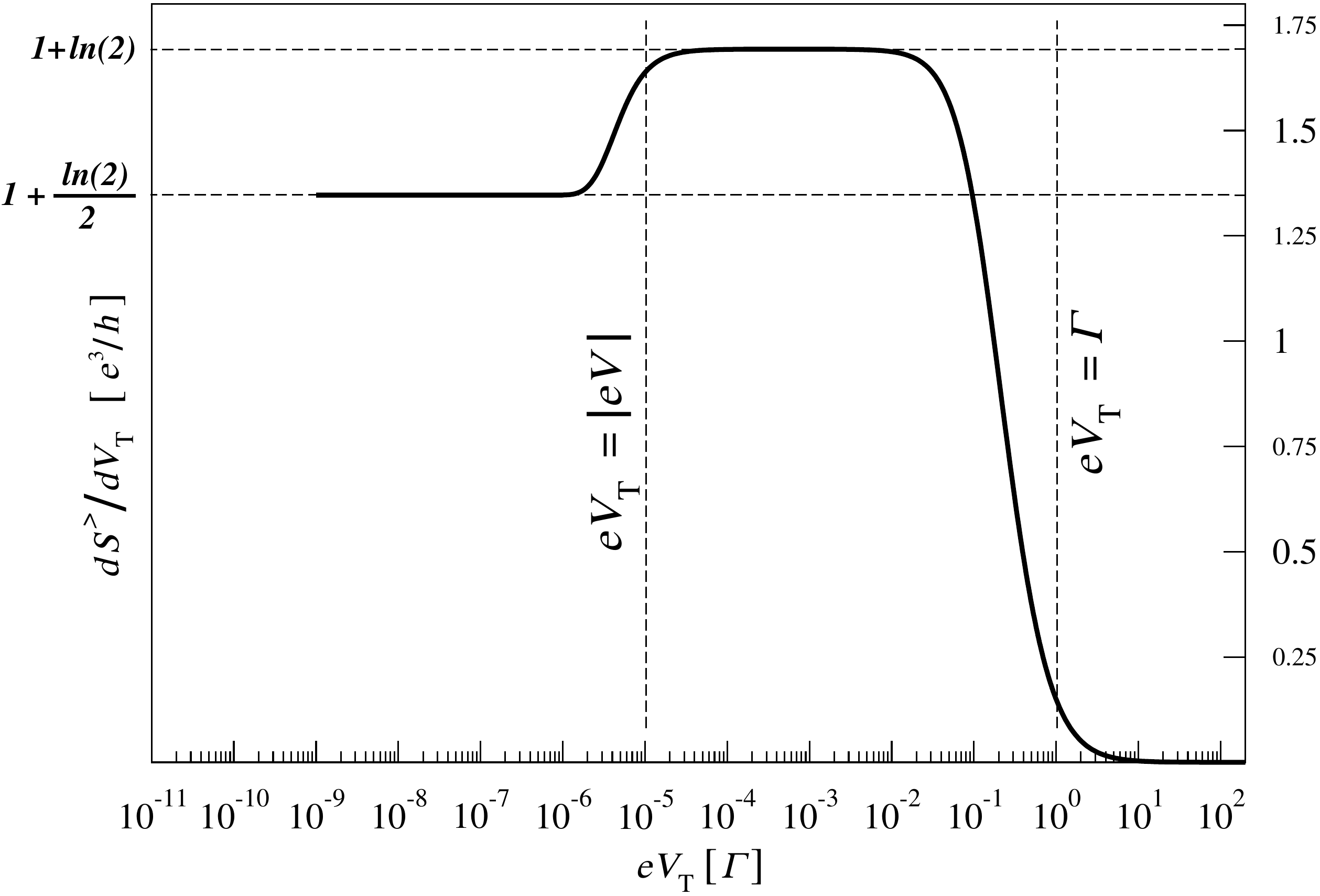}
\caption{\label{figure_6} The first derivative of the greater current-current
  correlator with respect to the thermal voltage,
  $\partial S^>(V,V_T)/\partial V_T$, as a universal (independent of
  $\epsilon_d$) function of the thermal voltage $V_T$ at $T=0$,
  $|eV|/\Gamma=10^{-5}$. The Majorana overlap energy and the strength of the
  Majorana tunneling are $\xi/\Gamma=10^{-4}$ and $|\eta|/\Gamma=10^3$,
  respectively.}
\end{figure}

The universal Majorana behavior of the fluctuation thermoelectric coefficient
$\partial S^>(V,V_T)/\partial V_T$ as a function of the bias voltage $V$ is
shown in Fig. \ref{figure_5} at zero temperature and $eV_T\ll\Gamma$. The
values of $\xi$ and $|\eta|$ are the same as for the mean electric current
from the previous subsection so that the Majorana overlap is very weak and the
Majorana tunneling is very strong. For
\begin{equation}
|\eta|\gg\Gamma\gg eV_T\gg |eV|,\quad\Gamma\gg\xi
\label{St_1_cond}
\end{equation}
it turns out that the fluctuation thermoelectric coefficient
$\partial S^>(V,V_T)/\partial V_T$ does not depend on $V$ and is truly
universal (independent of $\epsilon_d$, $\Gamma$ and $\eta$). The asymptotic
limit of this truly universal constant is
\begin{equation}
\frac{\partial S^>(V,V_T)}{\partial V_T}=\frac{e^3}{h}[1+\ln(2)].
\label{St_1}
\end{equation}
For
\begin{equation}
|\eta|\gg\Gamma\gg |eV|\gg eV_T,\quad\Gamma\gg\xi
\label{St_2_cond}
\end{equation}
the fluctuation thermoelectric coefficient $\partial S^>(V,V_T)/\partial V_T$
does not depend on $V$ either and is equal to a different truly universal
constant. The asymptotic limit in this case is
\begin{equation}
\frac{\partial S^>(V,V_T)}{\partial V_T}=\frac{e^3}{h}\biggl[1+\frac{1}{2}\ln(2)\biggl].
\label{St_2}
\end{equation}
These two plateaus are explicitly visible in Fig. \ref{figure_5}.
\begin{figure}
\includegraphics[width=8.0 cm]{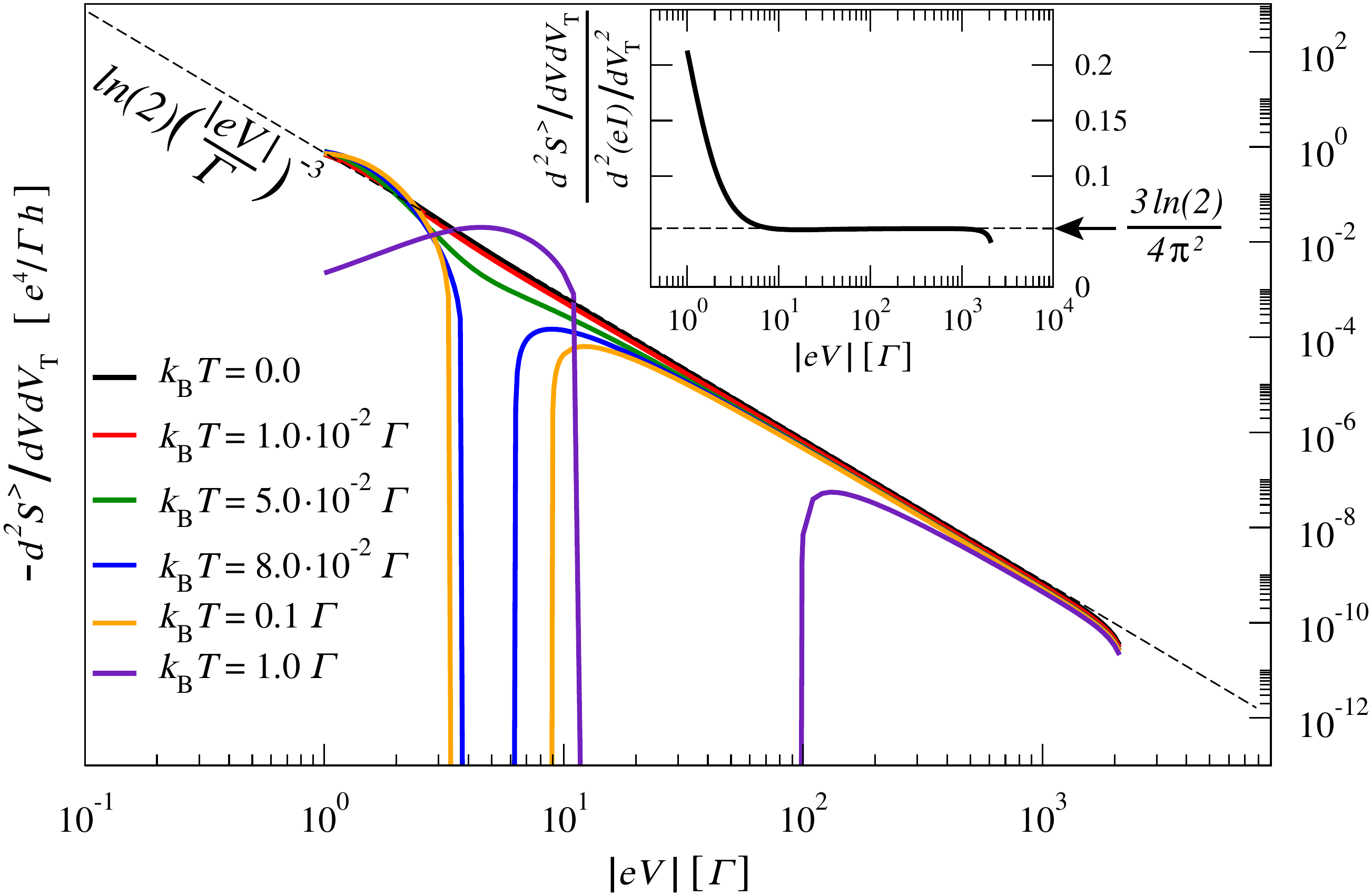}
\caption{\label{figure_7} The universal (independent of $\epsilon_d$) behavior
  of the second mixed derivative of the greater current-current correlator
  with respect to the thermal voltage and the bias voltage,
  $\partial^2S^>(V,V_T)/\partial V\partial V_T$, at large bias voltages
  $V$. The thermal voltage is $eV_T/\Gamma=10^{-2}$. The Majorana overlap
  energy and the strength of the Majorana tunneling are $\xi/\Gamma=10^{-4}$
  and $|\eta|/\Gamma=10^3$, respectively. The curves demonstrate robustness of
  $\partial^2S^>(V,V_T)/\partial V\partial V_T$ at $|eV|\gg k_\text{B}T$ when
  the temperature increases. The inset shows the universal nonlinear response
  thermoelectric ratio of the second mixed derivative of the greater
  current-current correlator with respect to the thermal voltage and the bias
  voltage to the second derivative of the mean electric current times the
  electron charge with respect to the thermal voltage,
  $\bigl\{\partial^2S^>(V,V_T)/\partial V\partial V_T\bigl\}/\bigl\{\partial^2[eI(V,V_T)]/\partial V_T^2\bigl\}$,
  at $T=0$.}
\end{figure}

In the high-energy regime specified in (\ref{It_2_cond}) the universal
asymptotic limit of the fluctuation thermoelectric coefficient
$\partial S^>(V,V_T)/\partial V_T$ is
\begin{equation}
\frac{\partial S^>(V,V_T)}{\partial V_T}=\frac{e^3}{h}\frac{\ln(2)}{2}\biggl(\frac{|eV|}{\Gamma}\biggl)^{-2}.
\label{St_3}
\end{equation}

As a function of the thermal voltage $V_T$ the fluctuation thermoelectric
\begin{figure}
\includegraphics[width=8.0 cm]{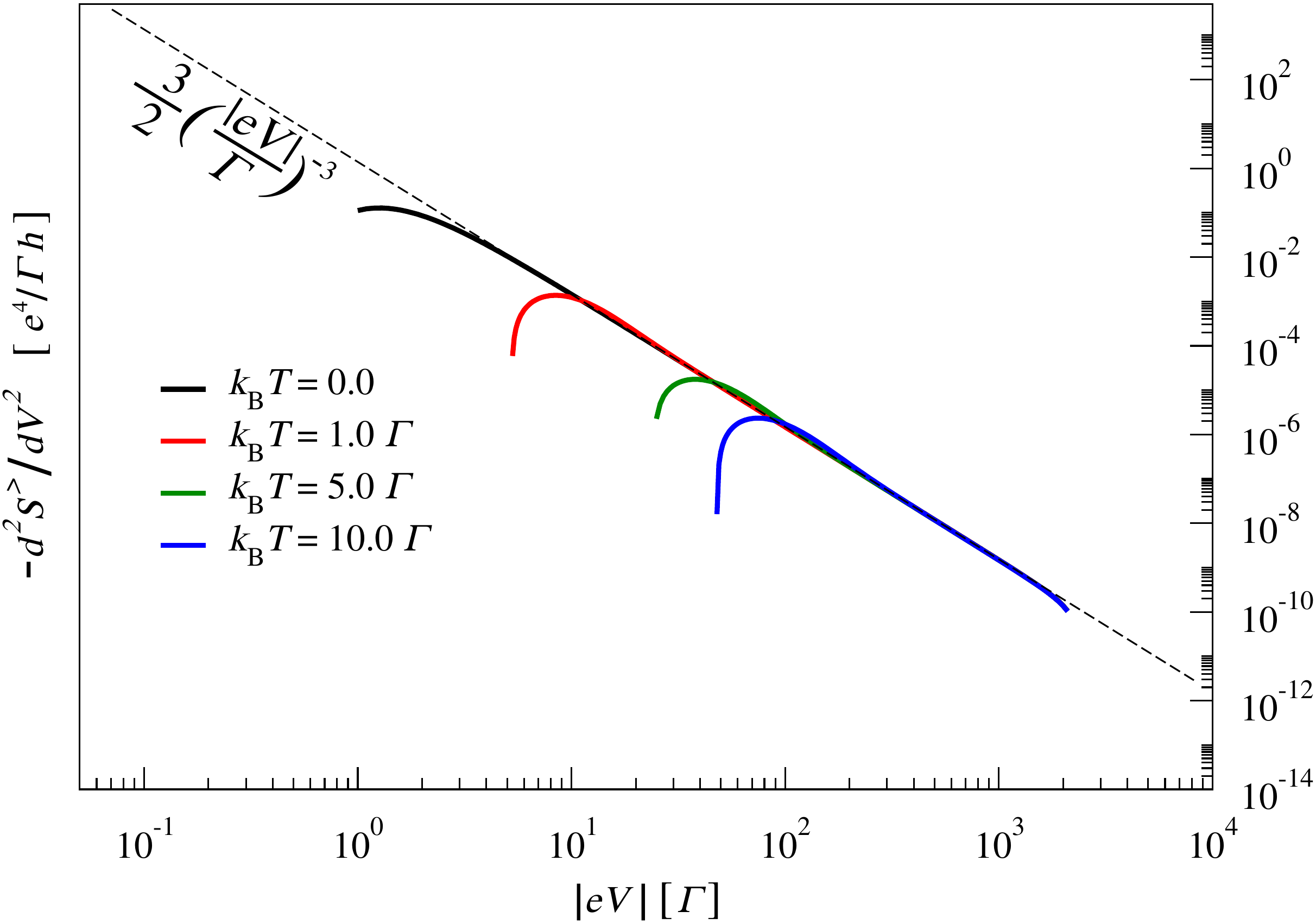}
\caption{\label{figure_8} The universal (independent of $\epsilon_d$) behavior
  of the second derivative of the greater current-current correlator with
  respect to the bias voltage, $\partial^2S^>(V,V_T)/\partial V^2$, at large
  bias voltages $V$. The thermal voltage is $eV_T/\Gamma=10^{-5}$. The
  Majorana overlap energy and the strength of the Majorana tunneling are
  $\xi/\Gamma=10^{-4}$ and $|\eta|/\Gamma=10^3$, respectively. The curves are
  shown for different temperatures and demonstrate remarkable robustness of
  $\partial^2S^>(V,V_T)/\partial V^2$ at $|eV|\gg k_\text{B}T$ when the
  temperature increases.}
\end{figure}
\begin{figure}
\includegraphics[width=8.0 cm]{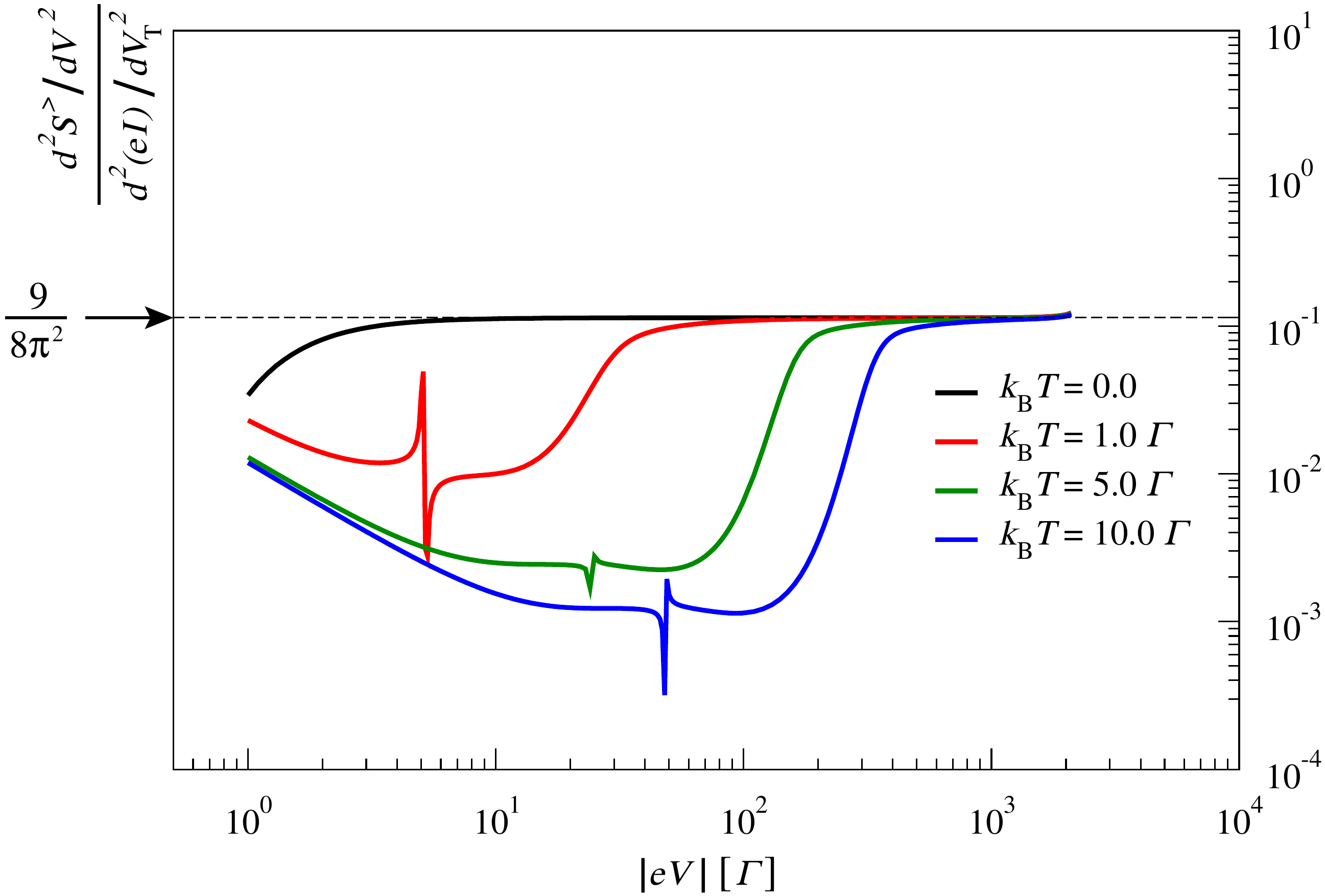}
\caption{\label{figure_9} The universal nonlinear response thermoelectric
  ratio of the second derivative of the greater current-current correlator
  with respect to the bias voltage to the second derivative of the mean
  electric current times the electron charge with respect to the thermal
  voltage,
  $\bigl\{\partial^2S^>(V,V_T)/\partial V^2\bigl\}/\bigl\{\partial^2[eI(V,V_T)]/\partial V_T^2\bigl\}$.
  The thermal voltage is $eV_T/\Gamma=10^{-5}$. The Majorana overlap energy
  and the strength of the Majorana tunneling are $\xi/\Gamma=10^{-4}$ and
  $|\eta|/\Gamma=10^3$, respectively. The curves are shown for different
  temperatures and demonstrate that at $|eV|\gg k_\text{B}T$ this universal
  nonlinear response thermoelectric ratio is remarkably robust against thermal
  fluctuations whose amplitude grows when the temperature increases.}
\end{figure}
coefficient $\partial S^>(V,V_T)/\partial V_T$ is shown in
Fig. \ref{figure_6}. In the regimes specified in (\ref{St_1_cond}) and
(\ref{St_2_cond}) this dependence has, respectively, the plateaus (\ref{St_1})
and (\ref{St_2}) with the only difference in the order of these plateaus. Here
the lowest plateau (\ref{St_2}) is followed by the highest one (\ref{St_1}).

The physical mechanism underlying the fact that the truly universal plateau
(\ref{St_1}) in the regime (\ref{St_1_cond}) is larger than the truly
universal plateau (\ref{St_2}) in the regime (\ref{St_2_cond}) can be
understood as follows. As schematically shown in Fig. \ref{figure_1}, at $T=0$
the Fermi-Dirac distribution in the left (hot) contact is smeared over the
energy range $eV_T$ around the Fermi energy of this contact. When we have
$eV_T\gg|eV|$, there exist two electron flows. The first flow, from the right
(cold) contact to the left contact, is induced by electrons coming from the
filled states below the Fermi energy of the right contact to the partially
filled states below and above the Fermi energy of the left contact. The second
flow, from the left contact to the right contact, is induced by electrons
coming from the partially filled states above the Fermi energy of the left
contact to the empty states above the Fermi energy of the right
contact. Both of these electron flows fluctuate and contribute to the total
noise $S^>(V,V_T)$ which grows linearly as a function of $V_T$ with the slope
given by Eq. (\ref{St_1}). However, when we decrease the thermal voltage up to
the bias voltage, $eV_T\sim|eV|$, the second flow significantly decays and
becomes negligible for $eV_T\ll|eV|$ because the population of the states in
the left contact above its Fermi energy rapidly goes to zero. In this
situation the noise contribution from the second flow vanishes and only
fluctuations of the first flow essentially contribute to the total noise
$S^>(V,V_T)$ which still grows linearly as a function of $V_T$ but with the
reduced slope given by Eq. (\ref{St_2}).

In the high-energy regime,
\begin{equation}
|\eta|\gg eV_T\gg\Gamma\gg |eV|,\,\,\,\Gamma\gg\xi,
\label{S_t_4_cond}
\end{equation}
however, we were unable to identify the universal asymptotic limit of the
fluctuation thermoelectric coefficient $\partial S^>(V,V_T)/\partial
V_T$. This high-energy universal law is definitely not a power dependence and
it will be a challenge for future research.

In Fig. \ref{figure_7} we show the high-energy universal behavior of the
fluctuation thermoelectric coefficient representing the second mixed
derivative of the greater noise with respect to the thermal voltage and the
bias voltage, $\partial^2 S^>(V,V_T)/\partial V\partial V_T$, as a function of
the bias voltage at $eV_T\ll\Gamma$ and for different temperatures. As one can
see, at $|eV|\gg k_\text{B}T$ the high-energy behavior of the fluctuation
thermoelectric coefficient $\partial^2 S^>(V,V_T)/\partial V\partial V_T$ is
almost independent of the temperature and obeys the universal asymptotic
limit,
\begin{equation}
\frac{\partial^2 S^>(V,V_T)}{\partial V\partial V_T}=-\frac{e^4}{\Gamma h}\ln(2)\biggl(\frac{|eV|}{\Gamma}\biggl)^{-3},
\label{S_vt}
\end{equation}
shown in Fig. \ref{figure_7} as the dashed line. The high-energy universal
behavior (\ref{S_vt}) follows from the high-energy universal behavior
(\ref{St_3}) of the fluctuation thermoelectric coefficient
$\partial S^>(V,V_T)/\partial V_T$.

Comparison of the high-energy universal behavior of the mean thermoelectric
coefficient $\partial^2I(V,V_T)/\partial V_T^2$ in Eq. (\ref{I_tt_2}) with the
high-energy universal behavior of the fluctuation thermoelectric coefficient
$\partial^2S^>(V,V_T)/\partial V\partial V_T$ in Eq. (\ref{S_vt}) shows that
the ratio between these two nonlinear response coefficients is a truly
universal constant with the asymptotic limit,
\begin{equation}
\frac{\frac{\partial^2S^>(V,V_T)}{\partial V\partial V_T}}{\frac{\partial^2[eI(V,V_T)]}{\partial V_T^2}}=
\frac{3\ln(2)}{4\pi^2},
\label{Univ_ratio_1}
\end{equation}
shown in the inset in Fig. \ref{figure_7} as the dashed horizontal line. As
can be seen in the inset, indeed, at zero temperature and large bias voltages,
$|eV|\gg\Gamma$, in the regime specified in (\ref{M_tunn_reg}) the
thermoelectric ratio between $\partial^2S^>(V,V_T)/\partial V\partial V_T$ and
$\partial^2[eI(V,V_T)]/\partial V_T^2$ is independent of $\epsilon_d$,
$\Gamma$, $\eta$ and the bias voltage $V$. However, outside the high-energy
regime, {\it e.g.}, $|eV|\sim\Gamma$, or when the Majorana tunneling condition
(\ref{M_tunn_reg}) is not satisfied, {\it e.g.}, $|eV|>|\eta|$, one can see
deviations from the truly universal value (\ref{Univ_ratio_1}). The inset in
Fig. \ref{figure_7} demonstrates that at low bias voltages there are strong
deviations from the truly universal value (\ref{Univ_ratio_1}). Since these
deviations depend on $\Gamma$, the behavior at low bias voltages is not truly
universal. Nevertheless the thermoelectric ratio between
$\partial^2S^>(V,V_T)/\partial V\partial V_T$ and
$\partial^2[eI(V,V_T)]/\partial V_T^2$ is still universal even at low bias
voltages because it does not depend on $\epsilon_d$ in this voltage range. As
expected, we find that outside the validity of the Majorana tunneling
condition (\ref{M_tunn_reg}), {\it e.g.}, $|eV|>|\eta|$, the thermoelectric
ratio between $\partial^2S^>(V,V_T)/\partial V\partial V_T$ and
$\partial^2[eI(V,V_T)]/\partial V_T^2$ is not universal ({\it i.e.} it depends
on $\epsilon_d$) because the Majorana tunneling is not effective at such large
bias voltages where the physics is not governed anymore by the Majorana
fermions of the topological superconductor.

In Fig. \ref{figure_8} we show the high-energy universal behavior of the
fluctuation nonlinear response coefficient representing the second derivative
of the greater noise with respect to the bias voltage,
$\partial^2S^>(V,V_T)/\partial V^2$, as a function of the bias voltage at
$eV_T\ll\Gamma$ and for different temperatures. In the regime specified in
(\ref{It_2_cond}) the fluctuation nonlinear response coefficient
$\partial^2S^>(V,V_T)/\partial V^2$ is pretty much insensitive at $|eV|\gg k_\text{B}T$
to the temperature increase and its universal asymptotic limit is
\begin{equation}
\frac{\partial^2S^>(V,V_T)}{\partial V^2}=-\frac{e^4}{\Gamma h}\frac{3}{2}\biggl(\frac{|eV|}{\Gamma}\biggl)^{-3}.
\label{S_vv}
\end{equation}
Comparing Eq. (\ref{I_tt_2}) with Eq. (\ref{S_vv}) one obtains that the
thermoelectric ratio between $\partial^2S^>(V,V_T)/\partial V^2$ and
$\partial^2[eI(V,V_T)]/\partial V_T^2$ is a truly universal constant with the
asymptotic limit
\begin{equation}
\frac{\frac{\partial^2S^>(V,V_T)}{\partial V^2}}{\frac{\partial^2[eI(V,V_T)]}{\partial V_T^2}}=\frac{9}{8\pi^2}
\label{Univ_ratio_2}
\end{equation}
in the regime specified in (\ref{It_2_cond}). This truly universal constant is
shown in Fig. \ref{figure_9} as the dashed horizontal line. As demonstrated in
Fig. \ref{figure_9}, in the regime specified in (\ref{It_2_cond}) the
universal Majorana thermoelectric ratio between the fluctuation nonlinear
response coefficient $\partial^2S^>(V,V_T)/\partial V^2$ and the mean
nonlinear coefficient $\partial^2[eI(V,V_T)]/\partial V_T^2$ is remarkably
robust at $|eV|\gg k_\text{B}T$ with respect to thermal noise excited at high
temperatures.

At this point it is important to notice that the truly universal ratios
(\ref{Univ_ratio_1}) and (\ref{Univ_ratio_2}), which are valid in the Majorana
tunneling regime (\ref{M_tunn_reg}) and at high bias voltages $|eV|\gg\Gamma$,
might result from symmetries of the full counting statistics as has been
discussed in Refs. \cite{Tobiska_2005,Andrieux_2006,Saito_2008,Iyoda_2010}. In
particular, in Refs. \cite{Saito_2008,Iyoda_2010} symmetries of the full
counting statistics were used to derive relations between nonlinear response
coefficients beyond the Onsager-Casimir relation. Generalization of those
relations between nonlinear response coefficients and derivation of
(\ref{Univ_ratio_1}) and (\ref{Univ_ratio_2}) using symmetries of the full
counting statistics in the presence of topological superconductors supporting
Majorana bound states is an important fundamental task which is, to our
knowledge, still unexplored and definitely represents a special topic for
future research.

Let us estimate at which temperatures one can experimentally observe the
truly universal high-energy Majorana plateau (\ref{Univ_ratio_2}) in the
thermoelectric ratio between the fluctuation nonlinear response coefficient
$\partial^2S^>(V,V_T)/\partial V^2$ and the mean nonlinear response
coefficient $\partial^2I(V,V_T)/\partial V_T^2$. We take the highest
temperature shown in Fig. \ref{figure_9}, that is $k_\text{B}T=10\,\Gamma$. In
units of $|\eta|$ we have $k_\text{B}T=0.01\,|\eta|$. Since $|\eta|$ is the
largest energy scale, it should not exceed the induced superconducting energy
gap $\Delta$ so as not to excite the bulk quasiparticles. Therefore, we assume
$|\eta|\sim\Delta$. In Ref. \cite{Mourik_2012} the induced superconducting
energy gap is estimated as $\Delta\approx 250\,\mu\text{eV}$. One then obtains
the temperature
$T\approx 2.5\,\mu\text{eV}/k_\text{B}\approx 0.03\,\text{K}=30\,\text{mK}$.
In Ref. \cite{Wang_2013} a higher value of $\Delta$ is reported,
$\Delta\approx 15\,\text{meV}$. In this case one has $T\approx 1.8\,\text{K}$.
Such temperatures are already high enough and thus may easily be achieved in
modern laboratories. As a consequence, the truly universal thermoelectric
constant (\ref{Univ_ratio_2}) represents a unique and highly conclusive Majorana
signature to be detected in a realistic experiment.
\section{Conclusion}\label{concl}
In this work we have explored thermoelectric Majorana response of both the
mean value and the fluctuations of the electric current. The research has been
focused on unique universal Majorana thermoelectric signatures which may be
detected in nanoscopic systems such as quantum dots. It has been shown that
mean thermoelectric quantities become universal only at bias voltages high
enough to exceed the overlap energy of the two Majorana bound states. For bias
voltages below the Majorana overlap energy mean thermoelectric quantities are
not universal even if the Majorana fermions are well separated. As a
consequence, in order to obtain universal Majorana signatures from
measurements of mean thermoelectric quantities one has to resort to
essentially nonlinear response. In contrast, the thermoelectric response of
the fluctuations of the electric current turns out to be universal at any bias
voltage if the Majorana fermions are well separated. We have obtained various
fluctuation thermoelectric coefficients in different transport regimes. In
particular, it has been shown that the differential thermoelectric noise has a
universal two plateau structure with the truly universal values of the
plateaus $(e^3/h)[1+\ln(2)]$ and $(e^3/h)[1+\ln(2^{1/2})]$ depending on
whether the ratio between the bias voltage and the thermal voltage
(characterizing the temperature difference between the contacts) is less than
or greater than one. Further, universal high-energy behavior of the
fluctuation nonlinear response coefficients has been presented and universal
ratios between nonlinear response coefficients of the thermoelectric noise and
the mean current have been obtained. Finally, we have demonstrated that at
large bias voltages these thermoelectric ratios saturate to truly universal
constants independent of the bias voltage and found the asymptotic limits of
these truly universal constants. Importantly, these truly universal constants
are protected by high bias voltages, making them robust against thermal
noise. This robustness is crucial for realistic measurements. Therefore, the
truly universal constants characterizing thermoelectric ratios at high bias
voltages represent unique truly universal Majorana signatures challenging
modern experiments on thermoelectric noise in quantum dots.

It is fair to mention that many issues within Majorana noise still remain to
be addressed. In particular, the above results have been demonstrated for a
simple model where only one single-particle energy level of the quantum dot is
involved in transport. However, in many realistic systems several
single-particle energy levels of the quantum dot may often contribute to
transport. It is, therefore, interesting whether the results demonstrated here
will change and, if so, what kind of Majorana fluctuation fingerprints one
might expect in a multilevel system. This question as well as many other
interesting issues will be a challenge for our future research on Majorana
noise.

\section*{Acknowledgments}
The author thanks Milena Grifoni, Wataru Izumida and Meydi Ferrier for
important discussions.

\appendix*

\section{Basic steps in the calculation of the current-current correlator}
Taking the second derivative in Eq. (\ref{Curr_curr_corr}) one easily finds
that the current-current correlator is the sum of a one-particle,
$S_{1_{ll'}}(t,t')$, and a two-particle, $S_{2_{ll'}}(t,t')$, terms,
\begin{equation}
\langle I_l(t)I_{l'}(t')\rangle=S_{1_{ll'}}(t,t')+S_{2_{ll'}}(t,t').
\label{Curr_curr_corr_sum_1_2}
\end{equation}

The one-particle contribution has the form:
\begin{widetext}
\begin{equation}
\begin{split}
&S_{1_{ll'}}(t,t')=\delta_{ll'}|T_l|^2(-\mathrm{i})\biggl(\frac{e}{2\hbar}\biggl)^2
\sum_k\bigl\{\bigl\langle[\psi_1(t')+\psi_2(t')][-\bar{\psi}_1(t)+\bar{\psi}_2(t)]\bigl\rangle_{S_K}
\bigl[G_{lk}^R(t-t')+G_{lk}^K(t-t')-\\
&-G_{lk}^A(t-t')\bigl]+
\bigl\langle[\psi_1(t)-\psi_2(t)][\bar{\psi}_1(t')+\bar{\psi}_2(t')]\bigl\rangle_{S_K}\bigl[-G_{lk}^R(t'-t)+G_{lk}^K(t'-t)+G_{lk}^A(t'-t)\bigl]\bigl\},
\end{split}
\label{Curr_curr_corr_1}
\end{equation}
\end{widetext}
where the angular brackets are defined in Eq. (\ref{Aver_zero_scr_fld}),
$G_{lk}^{R,A,K}(t-t')$ are, respectively, the retarded, advanced and Keldysh
Green's functions of the isolated contacts and we have performed the Keldysh
rotation of the Grassmann fields of the quantum dot,
\begin{widetext}
\begin{equation}
\begin{split}
&\psi_q(t)=\frac{1}{\sqrt{2}}\bigl[\psi_1(t)+q\psi_2(t)\bigl],\quad \bar{\psi}_q(t)=\frac{1}{\sqrt{2}}\bigl[\bar{\psi}_2(t)+q\bar{\psi}_1(t)\bigl].
\end{split}
\label{Keld_rot_qd}
\end{equation}
\end{widetext}
The one-particle contribution in Eq. (\ref{Curr_curr_corr_1}) involves
averages of products of only two Grassmann fields of the quantum dot. This
contribution contains only normal terms, that is terms of the form
$\langle\psi_s(t)\bar{\psi}_{s'}(t')\rangle_{S_K}$, where $s,s'=1,2$. It is
obvious that anomalous terms, that is terms of the form
$\langle\psi_s(t)\psi_{s'}(t')\rangle_{S_K}$ or
$\langle\bar{\psi}_s(t)\bar{\psi}_{s'}(t')\rangle_{S_K}$, cannot appear in
this case. However, such terms do arise when one averages products of four
Grassmann fields in the two-particle contribution $S_{2_{ll'}}(t,t')$.

The two-particle contribution has the form:
\begin{widetext}
\begin{equation}
\begin{split}
&S_{2_{ll'}}(t,t')=|T_l|^2|T_{l'}|^2\frac{(-1)}{\hbar^2}\biggl(\frac{e}{2\hbar}\biggl)^2\int_{-\infty}^{\infty}dt_1\int_{-\infty}^{\infty}dt_2\sum_{k_1,k_2}
\biggl\langle\bigl\{\bigl[-\psi_1(t)+\psi_2(t)\bigl]
\bigl[\bar{\psi}_1(t_2)G_{lk_2}^R(t_2-t)-\\
&-\bar{\psi}_1(t_2)G_{lk_2}^K(t_2-t)-\bar{\psi}_2(t_2)G_{lk_2}^A(t_2-t)\bigl]-
\bigl[G_{lk_2}^R(t-t_2)\psi_1(t_2)+G_{lk_2}^K(t-t_2)\psi_2(t_2)-\\
&-G_{lk_2}^A(t-t_2)\psi_2(t_2)\bigl]\bigl[-\bar{\psi}_1(t)+\bar{\psi}_2(t)\bigl]\bigl\}
\bigl\{\bigl[\psi_1(t')+\psi_2(t')\bigl]\bigl[\bar{\psi}_1(t_1)G_{l'k_1}^R(t_1-t')+\bar{\psi}_1(t_1)G_{l'k_1}^K(t_1-t')+\\
&+\bar{\psi}_2(t_1)G_{l'k_1}^A(t_1-t')\bigl]-\bigl[G_{l'k_1}^R(t'-t_1)\psi_1(t_1)+G_{l'k_1}^K(t'-t_1)\psi_2(t_1)+\\
&+G_{l'k_1}^A(t'-t_1)\psi_2(t_1)\bigl]\bigl[\bar{\psi}_1(t')+\bar{\psi}_2(t')\bigl]\bigl\}\biggl\rangle_{S_K}.
\end{split}
\label{Curr_curr_corr_2}
\end{equation}
\end{widetext}
Since the total Keldysh action in Eq. (\ref{Tot_Keld_act_zero_scr_fld}) is
quadratic, the average of four Grassmann fields is given as the sum of
products of the averages of two Grassmann fields,
\begin{equation}
\begin{split}
&\langle\psi_1\bar{\psi}_2\psi_3\bar{\psi}_4\rangle_{S_K}=\langle\psi_1\bar{\psi}_2\rangle_{S_K}\langle\psi_3\bar{\psi}_4\rangle_{S_K}-\\
&-\langle\psi_1\bar{\psi}_4\rangle_{S_K}\langle\psi_3\bar{\psi}_2\rangle_{S_K}-\langle\psi_1\psi_3\rangle_{S_K}\langle\bar{\psi}_2\bar{\psi}_4\rangle_{S_K}.
\end{split}
\label{Avr_4_Grass_fld}
\end{equation}
In Eq. (\ref{Avr_4_Grass_fld}) $\psi_1$, $\bar{\psi}_2$, $\psi_3$,
$\bar{\psi}_4$ schematically denote those Grassmann fields which are taken
from a given product of four square brackets in Eq. (\ref{Curr_curr_corr_2})
in the same order as these Grassmann fields appear in these four square
brackets.

All those terms in Eq. (\ref{Curr_curr_corr_2}) which correspond to the first
term in the right hand side of Eq. (\ref{Avr_4_Grass_fld}) do not represent
interest for the calculation of the greater current-current correlator
$S^>(t,t';V,V_T)$ in Eq. (\ref{S_grt}) because they give just the square of
the mean electric current which is subtracted from the current-current
correlator $\langle I_L(t)I_L(t')\rangle$ in Eq. (\ref{S_grt}).

All those terms in Eq. (\ref{Curr_curr_corr_2}) which correspond to the second
and third terms in the right hand side of Eq. (\ref{Avr_4_Grass_fld}) give,
respectively, the normal and anomalous two-particle contributions to
$S^>(t,t';V,V_T)$.

To treat the normal and anomalous contributions to $S^>(t,t';V,V_T)$ from both
$S_{1_{ll'}}(t,t')$ and $S_{2_{ll'}}(t,t')$ it is convenient to introduce
a particle-hole space via the Grassmann fields $\psi_{is}$ ($i=p,h$; $s=1,2$),
\begin{equation}
\begin{split}
&\psi_{is}(t)\equiv\bar{\psi}_s(t),\quad i=p,\\
&\psi_{is}(t)\equiv\psi_s(t),\quad i=h.
\end{split}
\label{P_h_space}
\end{equation}

The averages in Eqs. (\ref{Curr_curr_corr_1}) and (\ref{Curr_curr_corr_2}) are
then expressed in terms of the hole-particle, hole-hole and particle-particle
retarded, advanced and Keldysh Green's functions of the quantum dot via,
respectively, the following relations:
\begin{widetext}
\begin{equation}
\langle\psi_{hs}(t)\psi_{ps'}(t')\rangle_{S_K}=
\begin{pmatrix}
\mathrm{i}\,G_{hp}^R(t-t') & \mathrm{i}\,G_{hp}^K(t-t') \\
0 & \mathrm{i}\,G_{hp}^A(t-t')
\end{pmatrix},
\label{G_f_hp}
\end{equation}
\begin{equation}
\langle\psi_{hs}(t)\psi_{hs'}(t')\rangle_{S_K}=
\begin{pmatrix}
\mathrm{i}\,G_{hh}^K(t-t') & \mathrm{i}\,G_{hh}^R(t-t') \\
\mathrm{i}\,G_{hh}^A(t-t') & 0
\end{pmatrix},
\label{G_f_hh}
\end{equation}
\begin{equation}
\langle\psi_{ps}(t)\psi_{ps'}(t')\rangle_{S_K}=
\begin{pmatrix}
0 & \mathrm{i}\,G_{pp}^A(t-t') \\
\mathrm{i}\,G_{pp}^R(t-t') & \mathrm{i}\,G_{pp}^K(t-t')
\end{pmatrix}.
\label{G_f_pp}
\end{equation}
\end{widetext}

Since the total Keldysh action in Eq. (\ref{Tot_Keld_act_zero_scr_fld}) is
quadratic, the calculation of the Green's functions in the right hand sides of
Eqs. (\ref{G_f_hp})-(\ref{G_f_pp}) is reduced to calculations of the
corresponding elements of the inverse kernel of this action. This is a quite
simple, although somewhat lengthy, mathematical step which one may easily
perform. After this step one finds the expressions for the hole-particle,
hole-hole and particle-particle retarded, advanced and Keldysh Green's
functions of the quantum dot. All the retarded and advanced Green's functions
have already been found in Ref. \cite{Smirnov_2015} in the energy domain. For
completeness we give them here:
\begin{widetext}
\begin{equation}
\begin{split}
&G^R_{hp}(\epsilon)=\frac{N^R_{hp}(\epsilon)}{f(\epsilon)},\quad
G^R_{hh}(\epsilon)=\frac{-8\hbar(\eta^*)^2\epsilon}{f(\epsilon)},\quad G^R_{pp}(\epsilon)=\frac{-8\hbar\eta^2\epsilon}{f(\epsilon)},\\
&G^A_{hp}(\epsilon)=[G^R_{hp}(\epsilon)]^*,\quad G^A_{hh}(\epsilon)=[G^R_{pp}(\epsilon)]^*,\quad G^A_{pp}(\epsilon)=[G^R_{hh}(\epsilon)]^*,\\
&f(\epsilon)=4\epsilon^4-\epsilon^2(\Gamma^2+4\epsilon_d^2+4\xi^2+16|\eta|^2)+\xi^2(\Gamma^2+4\epsilon_d^2)+\mathrm{i}\,4\Gamma[\epsilon^3-\epsilon(\xi^2+2|\eta|^2)],\\
&N^R_{hp}(\epsilon)=2\hbar\{-4|\eta|^2\epsilon-(\xi^2-\epsilon^2)[\mathrm{i}\Gamma+2(\epsilon_d+\epsilon)]\}.
\end{split}
\label{R_A_G_f}
\end{equation}

Finally, for the Keldysh components of the hole-particle, hole-hole and
particle-particle Green's functions one obtains the following expressions:
\begin{equation}
\begin{split}
&G^K_{hp}(\epsilon)=\frac{N^K_{hp}(\epsilon)}{|f(\epsilon)|^2},\quad
G^K_{hh}(\epsilon)=\frac{N^K_{hh}(\epsilon)}{|f(\epsilon)|^2},\quad G^K_{pp}(\epsilon)=\frac{N^K_{pp}(\epsilon)}{|f(\epsilon)|^2},\\
&N^K_{hp}(\epsilon)=-2\,\mathrm{i}\,\Gamma\hbar\bigl\{(F_L(\epsilon)+F_R(\epsilon))(\xi^2-\epsilon^2)^2[\Gamma^2+4(\epsilon_d+\epsilon)^2]+\\
&+16|\eta|^2\epsilon[(F_L(\epsilon)+F_R(\epsilon))(\xi^2-\epsilon^2)(\epsilon_d+\epsilon)+|\eta|^2\epsilon(F_L(\epsilon)-F_L(-\epsilon)+F_R(\epsilon)-F_R(-\epsilon))]\bigl\},\\
&N^K_{hh}(\epsilon)=8\Gamma\hbar(\eta^*)^2\epsilon\bigl\{(\xi^2-\epsilon^2)[(F_L(\epsilon)+F_L(-\epsilon)+F_R(\epsilon)+F_R(-\epsilon))(\Gamma-2\,\mathrm{i}\,\epsilon_d)+\\
&+2\,\mathrm{i}\,\epsilon(F_L(-\epsilon)-F_L(\epsilon)+F_R(-\epsilon)-F_R(\epsilon))]+4\,\mathrm{i}\,|\eta|^2\epsilon(F_L(-\epsilon)-F_L(\epsilon)+F_R(-\epsilon)-F_R(\epsilon))\bigl\},\\
&N^K_{pp}(\epsilon)=-(N^K_{hh}(\epsilon))^*,
\end{split}
\label{K_G_f}
\end{equation}
\end{widetext}
where $F_{L,R}(\epsilon)\equiv 1-2f_{L,R}(\epsilon)$.

Using Eqs. (\ref{Curr_curr_corr_1}), (\ref{Curr_curr_corr_2}),
(\ref{Avr_4_Grass_fld}) and (\ref{G_f_hp})-(\ref{K_G_f}) one may without
much effort express the zero frequency noise in Eq. (\ref{S_gr_zr_fr}) as well
as its various derivatives as integrals in the energy domain. These integrals
may be calculated numerically to obtain the quantities discussed in the main
text as functions of the bias voltage and thermal voltage.

\end{document}